\documentclass[useAMS,usenatbib]{mn2e}
\pdfoutput=1

\usepackage{epsfig} 
\usepackage{multirow}

\renewcommand\({\left(}
\renewcommand\){\right)}
\renewcommand\[{\left[}
\renewcommand\]{\right]}

\newcommand{\ra}{\rightarrow}

\def\lsim{\raise 0.4ex\hbox{$<$}\kern -0.8em\lower 0.62
ex\hbox{$\sim$}}

\def\gsim{\raise 0.4ex\hbox{$>$}\kern -0.7em\lower 0.62
ex\hbox{$\sim$}}

\def\lbar{{\hbox{$\lambda$}\kern -0.7em\raise 0.6ex
\hbox{$-$}}}

\newcommand\eq[1]{eq.~(\ref{#1})}
\newcommand\eqs[2]{eqs.~(\ref{#1}) and (\ref{#2})}

\newcommand\Eqs[2]{Equations~(\ref{#1}) and (\ref{#2})}

\newcommand\pa{\partial}
\newcommand\p{\partial}

\newcommand\ee{\end{equation}}
\newcommand\be{\begin{equation}}
\def\bea{\begin{array}}
\def\eea{\end{array}}\def\ea{\end{array}}
\newcommand\ees{\end{eqnarray}}
\newcommand\bees{\begin{eqnarray}}

\def\p1{{\bf p}_1}
\def\p2{{\bf p}_2}
\def\k1{{\bf k}_1}
\def\k2{{\bf k}_2}

\newcommand{\dddM}{\kern 0.2em \raise 1.9ex\hbox{$...$}\kern -1.0em \hbox{$M$}}
\newcommand{\dddQ}{\kern 0.2em \raise 1.9ex\hbox{$...$}\kern -1.0em \hbox{$Q$}}
\newcommand{\dddI}{\kern 0.2em \raise 1.9ex\hbox{$...$}\kern -1.0em\hbox{$I$}}
\newcommand{\dddJ}{\kern 0.2em \raise 1.9ex\hbox{$...$}\kern-1.0em
\hbox{$J$}}
\newcommand{\dddcalJ}{\kern 0.2em \raise 1.9ex\hbox{$...$}\kern-1.0em
\hbox{${\cal J}$}}

\newcommand{\dddO}{\kern 0.2em \raise 1.9ex\hbox{$...$}\kern -1.0em
\hbox{${\cal O}$}}
\def\dddz{\raise 1.5ex\hbox{$...$}\kern -0.8em \hbox{$z$}}
\def\dddd{\raise 1.8ex\hbox{$...$}\kern -0.8em \hbox{$d$}}
\def\dddbd{\raise 1.8ex\hbox{$...$}\kern -0.8em \hbox{${\bf d}$}}
\def\ddbd{\raise 1.8ex\hbox{$..$}\kern -0.8em \hbox{${\bf d}$}}
\def\dddx{\raise 1.6ex\hbox{$...$}\kern -0.8em \hbox{$x$}}

\newcommand{\msun}{M_{\odot}}

\def\D{\Delta}
\def\p{\partial}
\def\a{\alpha}

\def\nn{\nonumber}

\def\s{\sigma}
\def\g{\gamma}

\def\d{\delta}

\def\eps{\epsilon}

\def\dslash{\hspace{-1mm}\not{\hbox{\kern-2pt $\partial$}}}
\def\Dslash{\not{\hbox{\kern-4pt $D$}}}
\def\pslash{\not{\hbox{\kern-2.1pt $p$}}}
\def\kslash{\not{\hbox{\kern-2.3pt $k$}}}
\def\qslash{\not{\hbox{\kern-2.3pt $q$}}}

\newcommand{\inT}{\int_{-\infty}^{\infty}}

\newcommand{\Dl}{\int{\cal D}\lambda}

\newcommand{\Bnp}{B_n^{(p)}}\newcommand{\Bnq}{B_n^{(q)}}

\title[Excursion Set Theory for   generic moving barriers and  non-Gaussian initial conditions ] 
{Excursion Set Theory 
for generic moving barriers and  non-Gaussian initial conditions}

\author[Andrea De Simone, Michele Maggiore and  Antonio Riotto]
{Andrea De Simone$^{1}$, Michele Maggiore$^{2}$ and  Antonio Riotto$^{3,4}$\\
\\
$^1$Center for Theoretical Physics and  Department of Physics,
Massachusetts Institute of Technology, Cambridge, MA 02139, USA\\
$^2$ D\'epartement de Physique Th\'eorique, 
Universit\'e de Gen\`eve, 24 quai Ansermet, CH-1211 Gen\`eve, Switzerland\\
$^3$ CERN, PH-TH Division, CH-1211, Gen\`eve 23,  Switzerland\\
$^4$INFN, Sezione di Padova, Via Marzolo 8,
I-35131 Padua, Italy
}

\begin{document}

\pagerange{ 
\pageref{firstpage}-- 
\pageref{lastpage}} 

\maketitle

\label{firstpage} 
\begin{abstract}
Excursion set theory, where density perturbations evolve stochastically with the smoothing scale,
provides a method for computing the mass function of cosmological structures like dark matter halos,  sheets and filaments. The computation of these mass functions is mapped into   the so-called first-passage time problem in the presence of a moving barrier. In this paper we  use the path integral formulation of the  excursion set theory developed recently to analytically solve  the first-passage time problem in the presence of a generic moving barrier, in particular the barrier corresponding to 
ellipsoidal collapse. We perform the computation for  both   Gaussian and non-Gaussian initial conditions and for a window function which is a top-hat in wavenumber space.
The expression of the halo mass function for  the ellipsoidal collapse barrier and with non-Gaussianity
is therefore obtained in a fully consistent way and    it does not require the introduction of any form factor artificially derived  from the Press-Schechter  formalism based on the spherical collapse and usually adopted in the literature.

\end{abstract}
\begin{keywords}
	cosmology: theory -- large scale structure of the universe
\end{keywords}

\section{Introduction}

The mass function of dark matter halos is a central object in modern cosmology,  because of its relevance to the formation and evolution of  galaxies and clusters. It is therefore important to have accurate theoretical predictions for it,  first of all when the primordial fluctuations are taken to be Gaussian, and then when some level of non-Gaussianity is included.
Non-Gaussianities  are particularly relevant in the  high-mass end of
the power spectrum of perturbations, i.e. on the scale of galaxy clusters,
since the effect of non-Gaussian (NG) fluctuations becomes especially visible on
the  tail of the probability distribution. 
As a result, both 
the abundance and  the clustering properties of very massive halos
are sensitive probes of primordial 
non-Gaussianities \citep{MLB,GW,LMV,MMLM,KOYAMA,MVJ,RB,RGS,LV,MR3,lam,porciani}, 
and could be detected or significantly constrained by
the various planned large-scale galaxy surveys,
both ground based (such as DES, PanSTARRS and LSST) and in space
(such as EUCLID and ADEPT) see,  e.g.  \cite{Dalal} and \cite{CVM}. 
Furthermore, the primordial NG
alters the clustering of dark matter halos inducing a scale-dependent
bias on large 
scales \citep{Dalal,MV,slosar,tolley}  while even for small primordial
NG the evolution of perturbations on super-Hubble scales yields
extra
contributions on 
smaller scales \citep{bartolosig,MV2009}.  

The formation and evolution of dark matter halos is a highly complex
phenomenon, and a detailed quantitative understanding of it can only come
through large-scale N-body simulations, such as the Millennium simulation
\citep{Springel:2005nw}. Simulations with non-Gaussian initial conditions have also been performed
\citep{grossi2009,porciani,t}. 
At the same time, some analytic understanding of the process of halo formation is also desirable, both for the deeper physical understanding that analytic models offer, and for their flexibility under changes of  parameters of the cosmological model, shape of non-Gaussianities, etc.
Analytical derivations of the halo mass function 
are typically based  on
Press-Schechter (PS) 
theory \citep{PS} and its extension~\citep{PH90,Bond} known as excursion set
theory	(see \cite{Zentner} for a recent review).
In  excursion set theory the density
perturbation evolves stochastically 
with the smoothing scale,
and the problem of computing the probability of halo formation is
mapped into the so-called first-passage time problem in the presence
of a barrier. 

The original formulation of excursion set theory \citep{Bond} makes a number of simplifying assumptions, both at the technical level, and concerning the physics of halo formation.
In particular, at the technical level it is assumed that 
the smoothed density field $\d$ evolves
with the smoothing scale $R$ (or more precisely with the variance
$S(R)$ of the smoothed density field) in a Markovian way.  However, this assumption is correct only if the density field is smoothed with a window function which is a top-hat in wavenumber space, and with such a smoothing function it is  difficult to associate a mass $M$ to a region smoothed with smoothing parameter $R$, so in practice it is not possible to associate a mass to the dark matter halos identified in this way.
For any other choice of the window function (such as a top-hat in real space, for which the relation between the mass $M$ and the smoothing scale $R$ is trivially $M=(4/3)\pi R^3\bar{\rho}$, where 
$\bar{\rho}$ is the average density of the universe), the actual evolution of the smoothed density field with $R$ is non-Markovian.
At the physical level, the crucial simplifying assumption of the original formulation of excursion set theory is that dark matter halo forms through the spherical collapse of initial overdensities. However  the actual process of halo formation, as revealed by N-body simulations,  is much more complicated, and involves  smooth accretion, tidal interactions with the environment, as well as violent episodes of collisions with other halos, merging and fragmentation.

In  a recent series of papers \citep{MR1,MR2,MR3}  (hereafter MR1, MR2 and MR3, respectively), the original formulation of excursion set theory has been extended to deal with the non-Markovian effects which are induced either by  the use of  a realistic filter function, or by   non-Gaussianities in the primordial density field.
The basic idea is to reformulate the first-passage time problem in the presence of a barrier in terms of the computation of a path integral with a boundary (i.e. over a sum over all ``trajectories'' $\d(S)$ that always stay below the barrier), and then to use  standard
results  from quantum field theory and statistical mechanics to express  this path integral in terms of the connected correlators of the theory. A path-integral with boundaries of the kind that we obtain is however not a very common object even in quantum field theory or statistical mechanics, and in MR1 and MR3 we developed the technique for evaluating it perturbatively with respect to  the
non-Markovian and the  non-Gaussian  effects.
This provided first of all a rederivation of the results of excursion set theory which, from the mathematical point of view, is from first principles (for instance the absorbing barrier boundary condition, which in the original formulation was imposed by hand, comes out automatically in the formalism of MR1). Furthermore it allows us to  include, at least perturbatively, 
the effect of non-Markovianities and of non-Gaussianities. In particular, in 
MR3 we have shown how to include the effect of a non-vanishing bispectrum, while the case of a non-vanishing trispectrum was considered in \cite{MR4} (see also \cite{D'Amico:2010ta} for an approach to non-Gaussianities which combines our technique with the saddle  point method developed in
\cite{MVJ}).

Of course this extension of excursion set theory, even if it provides an improvement of the original formulation  from the mathematical point of view,   still shares the same physical limitations of the original formulations, as long as the same model for collapse is used. The model for collapse can be improved in different, complementary, ways.  A crucial step was taken by  \cite{SMT} who took into account the fact that actual halos are triaxial \citep{BBKS,BondMyers} and showed that 
an ellipsoidal collapse  model can be implemented, within the excursion set theory framework, by computing the first-crossing rate in the presence of a barrier $B_{\rm el}(S)$ which depends on $S$ (``moving barrier"), rather than being constant at the value $\d_c$ of the spherical collapse,

\be\label{B(S)}
B_{\rm el}(S)\simeq \delta_c\[ 1 +0.4\(\frac{S}{\delta_c^2}\)^{0.6}\]\, .
\ee
Physically this reflects the fact  that low-mass halos
(which corresponds to  large $S$) have larger deviations from
sphericity and significant shear, that opposes  collapse. Therefore
low-mass halos require a higher density to collapse. In contrast,
very large halos are more and more spherical, so their
effective barrier reduces to the one for spherical collapse.
 In order to improve the agreement between the prediction from the
excursion set theory with an ellipsoidal collapse and the N-body 
simulations, \cite{SMT} also found that
it was necessary to replace $\d_c$ with $\sqrt{a}\d_c$, where  $\sqrt{a}\simeq 0.84$ was obtained
by requiring that their mass function 
fits  the GIF simulation. The moving barrier therefore becomes 

\be
\label{B(S)2}
B_{\rm el}(S)\simeq \sqrt{a}\,\, 
\delta_c\[ 1 +0.4\(\frac{S}{a\,\delta^2_c}\)^{0.6}\]\, .
\ee
The parameter  $a$  cannot be derived 
 from the dynamics of the ellipsoidal collapse. Rather on the
contrary,  the ellipsoidal collapse model would 
predict $a=1$ because in the  limit $S\equiv\s^2\ra 0$ (i.e. in the large mass limit) halos become more and more spherical, and therefore
the barrier must reduce to that of spherical  collapse. This mismatch might be originated by the fact that, 
as mentioned above, halo collapse is a very complex dynamical phenomenon, and modeling it as spherical, or even as ellipsoidal, is a significant  oversimplification. In addition, the very definition of 
what is a dark matter halo, both in N-body simulations and observationally, is a difficult problem.
In MR2 it was proposed  that some of the physical complications inherent to a realistic description of halo formation can be included in the excursion set theory framework, at least at an effective level, by taking into account that the critical value for collapse  is itself a stochastic variable, whose scatter reflects a number of complicated aspects 
of the underlying dynamics (see also \cite{Audit,LeeS,SMT} for earlier related ideas). Solving the first-passage time problem in the presence of a barrier which is diffusing around the value $\d_c$ of the spherical collapse model,
it was found in MR2 that the exponential factor in the Press-Schechter mass function changes from $\exp\{-\d_c^2/2\s^2\}$ to $\exp\{-a\d_c^2/2\s^2\}$, where
$a=1/(1+D_B)$ and $D_B$ is the diffusion coefficient of the barrier.  The numerical value of $D_B$, and therefore the corresponding value of $a$, depends among other things on the algorithm used for identifying halos. From recent N-body simulations that studied the properties of the collapse barrier, a value $D_B\simeq 0.25$ was deduced in MR2   predicting   $a\simeq 0.80$ (up to $\sigma$ smaller than about 3)  We remark that 
the deduced  value of $a$  also holds  when  the collapse is ellipsoidal which was a good fit to the
average threshold barrier found by N-body data. The value of $a\simeq 0.80$ 
isin excellent agreement with the exponential fall off of the mass function found in N-body simulations, for the same halo definition.

The path-integral formulation developed in MR1 and MR3 was restricted to the case of a constant barrier $\d_c$ (while in MR2 were considered  the stochastic fluctuations around it).
The aim of this paper is  to extend the path integral  formulation of excursion set theory to the case  of a   generic moving barrier, and to provide
analytical expressions which can be used to calculate  the corresponding 
first-passage time
probability.
 
Given  that the Sheth-Tormen (ST) halo mass function is widely used in the literature,
we believe that it is interesting to derive it by computing
the first-crossing rate
with an ellipsoidal barrier  
from first principles. To the best of our knowledge, an analytical expression of the first-crossing rate was given in \cite{ST2002} just
as a fit to the N-body data and its derivation has been sketched only recently in \cite{lam}. As we shall see, this derivation is not free from drawbacks. There are  other good reasons why solving analytically for the first-crossing rate with a generic moving barrier is interesting.  First,  excursion set theory can be   applied to characterize the cosmic web (\cite{shen}). 
 Combining models of triaxial collapse with  excursion set theory,  cosmic sheets are defined as objects
that have collapsed along only one axis, filaments have collapsed along two axes, and halos are objects in which
triaxial collapse is complete. Computing the abundances of cosmic sheets, filaments and halos within the excursion set theory amounts again to solving  a first-time passage problem with the corresponding  moving barriers

\begin{eqnarray}
\label{sf1}
B_{\rm sheet}(S)&\simeq& \sqrt{a}\,\, 
\delta_c\[ 1 -0.56\(\frac{S}{a\,\delta^2_c}\)^{0.55}\]\, ,\\
\label{sf2}
B_{\rm filam}(S)&\simeq& \sqrt{a}\,\, 
\delta_c\[ 1 -0.012\(\frac{S}{a\,\delta^2_c}\)^{0.28}\]\, .
\end{eqnarray}
The insertion of each moving barrier into the excursion set approach
provides estimates of the mass fraction in sheets, filaments and
halos as a function of mass and time. Secondly, moving barriers are adopted in modelling
through the excursion set method the sizes of ionized regions during the epoch of reonization (\cite{furl}), while \cite{ST2002} suggested that moving barriers could effectively incapsulate a wide variety of phenomena such as suppression of the collapse of small, low-mass, overdense patches in models
in which dark matter is warm. For a given choice of the barrier, the first-crossing rate can in principle be evaluated with numerical techniques \citep{Bond,Zhang:2005ar}, but it  interesting to obtain analytic formulas valid for a generic functions $B(S)$.
Thirdly, as we already mentioned, it has become recently clear that 
 detecting a
significant amount of 
non-Gaussianity  and its shape either from the Cosmic Microwave Background (CMB)  or from the
Large Scale Structure (LSS) offers the possibility of opening a   window  into the 
dynamics of the universe
during the very first 
stages of its evolution (\cite{bartoloreview}). It is therefore of primary importance to compute the halo mass function when NG initial conditions are present. The halo mass function with NG has been calculated in \cite{MVJ} and \cite{LV} using the PS approach with a spherical collapse, while  the path integral formulation of excursion set theory
in the presence of NG and with a diffusive barrier has been formulated in MR3. 
The main motivation to compute the halo mass function in the presence of NG 
 within the excursion set method and with a moving ellipsoidal barrier is dictated by the fact that it has become customary  in the 
 literature to obtain the halo mass function with NG  by multiplying the ST halo mass function
with gaussian initial conditions by a form factor  obtained
by dividing the first-crossing rate with  NG obtained for  the PS spherical collapse case (\cite{MVJ,LV})
by the PS one (the  exception is represented by the consistent calculation of MR3, which does not require this procedure).  It is unclear (at least to us) why and to which extent this spurious method should provide a good approximation to the correct halo mass function with NG and ellipsoidal barrier. The issue is also timely since N-body data with NG  initial conditions finally exist \citep{grossi2009,porciani,t}, and may be compared to the various theoretical predictions for the halo mass functions with NG. They differ  
at the  ${\cal O}(20)$\% level and it is important to understand which error is introduced by adopting the 
form factor procedure.

The paper is organized as follows.
In section~\ref{sect:halo} we   review the approach to the computation of the halo mass function based on excursion set theory. In particular, in section~\ref{sect:spher}
we  begin with a quick review  of the case in which the collapse is assumed to be spherical, primordial fluctuations are taken to be Gaussian, and the evolution of the density perturbation with the smoothing scale is assumed to be Markovian. This is the setting considered in the classical paper by \cite{Bond}. We will then proceed toward increasing complexity. In Section~\ref{sect:pi}  we  review the the basic points of the approach developed in MR1, MR2 and MR3. In 
Section~\ref{sect:ellips} 
we present the computation of the first crossing rate for a generic moving barrier, while 
Section~\ref{sect:ellNG} contains the generalization of the computation to the case of NG initial conditions. Various technical details  are collected in Appendices A-D.

\section{The halo mass function in excursion set theory}\label{sect:halo}

The halo mass function can be written as
\be
\label{dndMdef}
\frac{dn(M)}{dM} = f(\s) \frac{\bar{\rho}}{M^2} 
\frac{d\ln\s^{-1} (M)}{d\ln M}\, ,
\ee
where $n(M)$ is the  number density of dark matter halos of mass $M$,
$\s(M)$ is the variance of the linear density field smoothed on a
scale $R$ corresponding to a mass $M$, and
$\bar{\rho}$ is the average density of the universe. 
The basic problem is therefore the computation of the function $f(\s)$.

\subsection{Spherical collapse, Gaussian fluctuations, and Markovian evolution with the smoothing scale}\label{sect:spher}

Let us summarize the basic points of the original formulation of 
excursion set theory. One
considers the density field $\d$  smoothed over a radius $R$, and studies its stochastic evolution as a function
of the smoothing scale $R$.
As it was found in the classical paper by
\cite{Bond},
when the density $\d(R)$  is smoothed with a sharp filter in momentum
space, and the density fluctuations have Gaussian statistics, the smoothed
density field satisfies the equation
\be\label{Langevin1}
\frac{\pa\d(S)}{\pa S} = \eta(S)\, ,
\ee
where $S=\s^2(R)$ is the variance of the linear density field
smoothed on the scale $R$ and computed with a sharp filter in momentum
space, while
$\eta (S)$ is a stochastic variable that satisfies 
\be\label{Langevin2}
\langle \eta(S_1)\eta(S_2)\rangle =\d_D (S_1-S_2)\, ,
\ee
where $\d_D$ denotes the Dirac delta function.
\Eqs{Langevin1} {Langevin2} are   the same as a Langevin equation with a
Dirac-delta  noise $\eta(S)$, with the variance $S$ formally playing the role
of time. 
Let us denote by $\Pi(\d,S)d\d$ the probability density that 
the variable $\d(S)$ reaches a value between $\d$ and $\d+d\d$ by ``time" $S$. 
A textbook  result in statistical physics is that, if a variable $\d(S)$
satisfies a Langevin equation with a Dirac-delta  noise, the  probability
density
$\Pi(\d,S)$ satisfies the Fokker-Planck (FP) equation
\be\label{FPdS}
\frac{\pa\Pi}{\pa S}=\frac{1}{2}\, \frac{\pa^2\Pi}{\pa \d^2}\, .
\ee
The solution of this equation over the whole real axis 
$-\infty<\d<\infty$, with the boundary condition that it vanishes at
$\d=\pm\infty$, is
\be\label{singlegau}
\Pi^0 (\d,S)=\frac{1}{\sqrt{2\pi S}}\, e^{-\d^2/(2S)}\, .
\ee
and is nothing but the distribution function of PS theory. Since, in hierarchical models of structure formation, as $R$ increases,
i.e. as the  halo mass increases,  the variance $S$ decreases monotonically, in 
\cite{Bond} it was realized that we are  actually interested in the stochastic
evolution of $\d$ against $S$ only until the ``trajectory" crosses for the first
time  the threshold $\d_c$ for collapse. The threshold value $\delta_c$ is estimated within the
spherical collapse model where a spherically symmetric inhomogeneity
behaves like a closed collapsing universe. 
The underlying idea behind the PS theory is that 
 the comoving number density of collapsed haloes can  computed from the statistical properties of the linear density field, assumed to be Gaussian. In this picture haloes form when the smoothed linear
 density contrast is larger than $\d_c\simeq 1.68$ which is obtained computing the linear density
 contrast at the collapse time. This result can be extended to arbitrary redshift
$z$ by reabsorbing the evolution of the variance into $\d_c$, so that
$\d_c$ in the above result is replaced by $\d_c(z)=\d_c(0)/D(z)$, where
$D(z)$ is the linear growth factor.
Notice that all the subsequent stochastic
evolution of $\d$ as a function of $S$, which in general results in trajectories
going multiple times above and below the threshold, is irrelevant, since it
corresponds to smaller-scale structures that will be erased and
engulfed by the collapse and virialization of the halo corresponding to the
largest value of $R$, i.e. the smallest value of $S$, for which the threshold
has been crossed.  In other words, trajectories
should be eliminated from further consideration once they have
reached the threshold for the first time. In \cite{Bond} this is implemented by
imposing the boundary condition
\be\label{bc}
\left.\Pi (\d,S)\right|_{\d=\d_c}=0\, .
\ee
The solution of the FP equation with this boundary condition
is
\be\label{PiChandra}
\Pi (\d,S)=\frac{1}{\sqrt{2\pi S}}\,
\[  e^{-\d^2/(2S)}- e^{-(2\d_c-\d)^2/(2S)} \]
\, ,
\ee
and gives the distribution function of  excursion set theory. 
The first term is the PS result, while the second term in \eq {PiChandra} is 
an ``image" Gaussian centered in $\d=2\d_c$.  Integrating this
$\Pi (\d,S)$ over $d\d$ from $-\infty$ to $\d_c$ gives the probability that a
trajectory, at ``time" $S$, has always been below the threshold. Increasing  $S$
this integral decreases because more and more trajectories cross the threshold
for the first time, so the  probability of first
crossing the threshold between ``time'' $S$ and $S+dS$ is given by
${\cal F}(S) dS$, with
\be\label{firstcrossT}
{\cal F}(S)  = -\frac{\pa}{\pa S}
\int_{-\infty}^{\d_c}d\d\, \Pi(\d;S)\, .
\ee
With standard manipulations (see e.g. \cite {Zentner} or MR1) one then
finds that the function $f(\s)$ which appears in \eq {dndMdef} is given by
\be\label{fcalF}
f(\s)= 2\s^2{\cal F}(\s^2)\, ,
\ee
where we wrote $S=\s^2$. Using \eq {PiChandra} one finds the PS prediction for the function $f(\s)$,
\bees
f_{\rm PS}(\s) &=& 
\(\frac{2}{\pi}\)^{1/2}\, 
\frac{\d_c}{\s}\, 
\, e^{-\d_c^2/(2\s^2)}\nn\\
&=&\(\frac{2}{\pi}\)^{1/2}\, 
\frac{\d_c}{S^{1/2}}\, 
\, e^{-\d_c^2/(2S)}\, ,\label{fps}
\ees
Observe that, when computing the first-crossing rate, the contribution of the
Gaussian centered in $\d=0$
and of the image Gaussian in \eq {PiChandra} add up, giving the well-known
factor of two that was missed in the original PS theory.

\subsection{Path integral formulation of  excursion set theory}
\label{sect:pi}
While excursion set theory is quite elegant, and gives a first analytic
understanding of the halo mass function, it suffers of two important set of
problems. First, it is based on the spherical collapse model,
which is, as we already mentioned,  a significant oversimplification of the actual complex dynamics of
halo formation.   
The second set of problems of excursion set theory is of a more technical
nature, and is due to the fact that the Langevin equation with Dirac-delta
noise, which is at the basis of the whole construction, can only
be derived if one works with a sharp filter in {\em momentum} space, and if
the fluctuations are Gaussian. However, as it
is well known \citep{Bond}, and as we have discussed at length in 
MR1, with such a
filter it is difficult  to associate a halo mass to the smoothing scale $R$.
When one uses a sharp filter in coordinate space, the evolution of the density
with the smoothing scale  becomes non-Markovian, and  the corresponding first-passage time problem is technically much more difficult. In particular, the distribution
function $\Pi(\d,S)$ no longer satisfies a local differential equation such as
the FP equation. The issue is particularly relevant when one wants to include
non-Gaussianities in the formalism, since 
the inclusion of non-Gaussianities renders again the dynamics non-Markovian.
Neglecting the non-Markovian dynamics due to the filter function would lead to incorrectly assigning to 
 non-Gaussianities in the primordial density field effects which are rather due, more trivially, to the procedure that one has adopted for smoothing  the density field.

In MR1,MR3  has been developed a formalism that allows us to
generalize excursion set theory to the case of a non-Markovian dynamics, either
generated by the filter function or by primordial non-Gaussianities. The basic
idea is the following. Rather than trying to derive a simple, local,
differential equation for $\Pi(\d,S)$ (which, as shown in MR1,
is impossible; in the non-Markovian case $\Pi(\d,S)$  rather satisfies a very
complicated equation which is non-local with respect to ``time" $S$), we
construct the probability distribution $\Pi(\d,S)$ directly by summing over all
paths that never exceeded the threshold $\d_c$, i.e. 
by writing $\Pi(\d,S)$ as a
path integral with boundaries.	To obtain such a representation, 
we consider an ensemble of
trajectories all starting at $S_0=0$ from an initial position
$\d(0) =\d_0$ 
and we follow them for a ``time'' $S$.	
We discretize the interval $[0,S]$ in steps
$\D S=\eps$, so $S_k=k\eps$ with $k=1,\ldots n$, and $S_n\equiv S$. 
A trajectory is  then defined by
the collection of values $\{\d_1,\ldots ,\d_n\}$, such that $\d(S_k)=\d_k$.
The probability density in the space of  trajectories is 
\be\label{defW}
W(\d_0;\d_1,\ldots ,\d_n;S_n)\equiv \langle
\d_D (\d(S_1)-\d_1)\ldots \d_D (\d(S_n)-\d_n)\rangle\, ,
\ee
where  $\d_D$
denotes the Dirac delta. Then the probability of arriving in $\d_n$ in a
``time'' $S_n$, starting from an initial value $\d_0$, without ever
going above the threshold, is\footnote{In \eqs
{singlegau} {PiChandra} we had implicitly assumed $\d_0=0$. In the following
however it will be necessary to keep track also of the initial position $\d_0$.}
\bees\label{defPi}
\Pi_{\eps} (\d_0;\d_n;S_n)
& \equiv&\int_{-\infty}^{\d_c} d\d_1\ldots \int_{-\infty}^{\d_c}d\d_{n-1}\nn\\
&&\times W(\d_0;\d_1,\ldots ,\d_{n-1},\d_n;S_n).
\ees
The label $\eps$ in $\Pi_{\eps}$ reminds us that this quantity is defined with
a finite spacing $\eps$, and we are finally interested in the continuum limit
$\eps\ra 0$.
As discussed in MR1  and MR3 (see Eqs.  (23)-(27) and discussion therein), $W(\d_0;\d_1,\ldots ,\d_{n-1},\d_n;S_n)$ can
be expressed in terms of the connected correlators of the theory,
\be\label{WnNG}
W(\d_0;\d_1,\ldots ,\d_n;S_n)=\Dl \, e^Z\, ,
\ee
where
\be
\Dl \equiv
\inT\frac{d\lambda_1}{2\pi}\ldots\frac{d\lambda_n}{2\pi}\, ,
\ee
and
\bees
Z&=& i\sum_{i=1}^n\lambda_i\d_i\label{defZ}\\
&&+\sum_{p=2}^{\infty}
\frac{(-i)^p}{p!}\,
 \sum_{i_1=1}^n\ldots \sum_{i_p=1}^n
\lambda_{i_1}\ldots\lambda_{i_p}\,
\langle \d_{i_1}\ldots\d_{i_p}\rangle_c\, .\nn
\ees
We also used the notation
$\d_i=\d(S_i)$,
and $\langle \d_{1}\ldots\d_{n}\rangle_c$ denotes the connected $n$-point
correlator. So
\be\label{Piexplicit}
\Pi_{\eps}(\d_0;\d_n;S_n)=\int_{-\infty}^{\d_c} d\d_1\ldots d\d_{n-1}\,
\Dl\, e^Z\, .
\ee
When $\d(S)$ satisfies \eqs {Langevin1} {Langevin2} (which is the case
for sharp filter in wavenumber space) the two-point function can be easily
computed, and is given by
\be\label{twopoint}
\langle\delta(S_i)\delta(S_j)\rangle = {\rm min}(S_i,S_j)\, .
\ee
If furthermore we consider Gaussian fluctuations,  all $n$-point connected
correlators with $n\geq 3$ vanish, and 
the probability density $W$ can be computed explicitly,
\be\label{W}
W^{\rm gm}(\d_0;\d_1,\ldots ,\d_n;S_n)=\frac{1}{(2\pi\eps)^{n/2}}\, 
e^{-\frac{1}{2\eps}\,
\sum_{i=0}^{n-1}  (\d_{i+1}-\d_i)^2}\hspace{-2mm},
\ee
where the superscript ``gm'' (Gaussian-Markovian) reminds us that this value of
$W$ is computed for Gaussian fluctuations, and when the evolution with respect to the
smoothing scale is Markovian.
Using this result, in MR1 we have shown that, in the continuum
limit, the distribution function $\Pi_{\eps=0}(\d;S)$, computed
with a sharp filter in
wavenumber space, satisfies a Fokker-Planck equation with the boundary
condition $\Pi_{\eps=0}(\d_c,S)=0$, and we have therefore recovered, from a
path integral approach, the distribution function of excursion set
theory,  \eq {PiChandra}. 
Considering a more realistic   filter, such as a step function in coordinate space, necessarily introduces
non-Markovianity and the computation, which is quite
non-trivial from a technical point of view, has been discussed in great detail
in MR1. In order to make the computation of the first-crossing rate
with a moving barrier more clear, from now on we will adopt the step function in wavenumber space
as a filter and eliminate the source of non-Markovianity given by the choice of the window function. 
The effect of a more realistic filter function could then be computed as in MR1. 
The effect, however,
will be tiny and totally negligible in the large mass range we are mostly interested in for the
non-Gaussian case.
Let us just close this subsection by reminding the reader about some useful properties of the path integral formulation which will turn out to be useful in the following. We will encounter objects such as 
\be\label{expr}
\sum_{i=1}^{n-1}F(S_i)\int_{-\infty}^{\delta_c} 
d\delta_1\ldots d\delta_{n-1}\,\pa_iW^{\rm gm}(\delta_0;\delta_1,\ldots
,\delta_n;S_n)\, ,\nn
\ee
where $F$ denotes a generic function.
To compute this expression we  integrate $\pa_i$ by parts,
\bees
&&\int_{-\infty}^{\delta_c} 
d\delta_1\ldots d\delta_{n-1}\,\pa_iW^{\rm gm}(\delta_0;\delta_1,\ldots
,\delta_n;S_n)\nn\\
&=&\int_{-\infty}^{\delta_c} d\delta_1\ldots \widehat{d\d}_i\ldots 
d\delta_{n-1}\label{byparts}\\
&&\times W(\delta_0;\delta_1,\ldots, \delta_i=\delta_c,\ldots
,\delta_{n-1},\delta_n;S_n)\, ,\nn
\ees
where
the notation $ \widehat{d\d}_i$ means that we must omit $d\delta_i$ from
the list of integration variables. 
We next observe that $W^{\rm gm}$ satisfies
\bees\label{facto}
&&W^{\rm gm}(\delta_0;\delta_1,\ldots, \delta_{i}= \delta_c, 
\ldots ,\delta_n; S_n)\nn\\
&=& 
W^{\rm gm}(\delta_0;\delta_1,\ldots, \delta_{i-1}, \delta_c;S_i)\nn\\
&&\times W^{\rm gm}(\delta_c; \delta_{i+1}, \ldots ,\delta_n; S_n-S_i)\, ,
\ees
as can be verified directly from its explicit expression (\ref{W}). Then
\bees\label{WPiPi}
&&\hspace*{-5mm}\int_{-\infty}^{\delta_c}d\delta_1\ldots d\delta_{i-1}
\int_{-\infty}^{\delta_c}  d\delta_{i+1}\ldots	d\delta_{n-1}\nn\\
&&\hspace*{-5mm}\times W^{\rm gm}(\delta_0;\delta_1,\ldots, \delta_{i-1}, \delta_c;S_i)
W^{\rm gm}(\delta_c; \delta_{i+1}, \ldots ,\delta_n; S_n-S_i)\nn\\
&&\hspace*{-5mm}= \Pi^{\rm gm}_{\eps}(\delta_0;\delta_c;S_i)
\Pi^{\rm gm}_{\eps}(\delta_c;\delta_n;S_n-S_i)\, ,
\ees
and to compute the expression given in \eq{expr} we must compute objects such as
\be\label{Pimem1}
\sum_{i=1}^{n-1}F(S_i) \Pi^{\rm gm}_{\eps}(\delta_0;\delta_c;S_i)
\Pi^{\rm gm}_{\eps}(\delta_c;\delta_n;S_n-S_i) .
\ee
To proceed further, we need to know $\Pi^{\rm
gm}_{\eps}(\delta_0;\delta_c;S_i)$. By definition, for $\eps=0$ this quantity
vanishes, since its second argument is equal to the the threshold value $\d_c$,
compare with \eq{bc}. However, in the continuum limit the sum over $i$ becomes
$1/\eps$ times an integral over an intermediate time variable $S_i$,
\be\label{sumint}
\sum_{i=1}^{n-1}\ra \frac{1}{\eps}\int_o^{S_n} dS_i\, ,
\ee
so we need to know how $\Pi^{\rm gm}_{\eps}(\delta_0;\delta_c;S_i)$ approaches
zero when $\eps\ra 0$.	In MR1 we proved that it vanishes as
$\sqrt{\eps}$, and that
\be\label{Pigammafinal}
\Pi^{\rm gm}_{\eps} (\delta_0;\delta_c;S)=
\sqrt{\eps}\, \frac{1}{\sqrt{\pi}}\, 
\frac{\delta_c-\delta_0}{S^{3/2}} e^{-(\delta_c-\delta_0)^2/(2S)} +{\cal
O}(\eps)\, .
\ee
Similarly, for $\delta_n<\delta_c$,
\be\label{Pigammafinalbis}
\Pi^{\rm gm}_{\eps} (\delta_c;\delta_n;S)=
\sqrt{\eps}\, \frac{1}{\sqrt{\pi}}\, 
\frac{\delta_c-\delta_n}{S^{3/2}} e^{-(\delta_c-\delta_n)^2/(2S)} +{\cal
O}(\eps)\, .
\ee
{
In the following, we will also need the expression for $\Pi^{\rm gm}_{\eps}$ 
with the  first and second 
argument both equal to $\d_c$, which is given by (see again MR1)
\be\label{Pigammafinalter}
\Pi^{\rm gm}_{\eps} (\delta_c;\delta_c;S)=
{\eps\over \sqrt{2\pi} S^{3/2}}\, .
\ee 
}
The two factors $\sqrt{\eps}$ from 
\eqs{Pigammafinal}{Pigammafinalbis} produce just an 
overall factor of $\eps$ that
compensates the factor $1/\eps$ in \eq {sumint}, and we are left with a finite
integral over $dS_i$.  Terms with two or more derivative, e.g. $\pa_i\pa_j$, or
$\pa_i,\pa_j\pa_k$ acting on $W$, with all indices $i,j,k$ maller than $n$, can
be computed similarly, and  have been discussed in detail in
MR1. With these technical details in mind, one can proceed to the computation of the first-crossing rate
in the presence of a moving barrier.

\section{Path integral with moving barrier: Gaussian fluctuations and Markovian evolution with the smoothing scale}
\label{sect:ellips}
In this section we  discuss the first-crossing rate for a generic moving barrier $B(S)$, specializing to the ellipsoidal one at the end.  We consider first the case of Gaussian primordial fluctuations,
and we will assume  that the evolution with the smoothing scale is Markovian.
Similarly to the constant barrier case, the probability of arriving at $\d_n$ in a
``time'' $S_n$, starting from the initial value $\delta_0=0$, without ever
going above the threshold, is
\bees\label{defPimoving}
\Pi_{\eps} (\d_n;S_n)
& \equiv&\int_{-\infty}^{B(S_1)} d\d_1\ldots \int_{-\infty}^{B(S_{n-1})}d\d_{n-1}\, 
\\
&&\times W(\d_0;\d_1,\ldots ,\d_{n-1},\d_n;S_n).\nn
\ees
Since we are considering the Gaussian and Markovian case,
$W(\d_0;\d_1,\ldots ,\d_{n-1},\d_n;S_n)$ can
be expressed in terms of the connected two-point function of the theory, as
\bees
&&W(\d_0;\d_1,\ldots ,\d_n;S_n)= \Dl\nn\\
&&\times\exp\left\{ i\sum_{i=1}^n\lambda_i\d_i-\frac{1}{2}\sum_{i,j=1}^n
\lambda_{i}\lambda_{j}\,
{\rm min}(S_i,S_j)
\right\}\, .
\ees
Taking the derivative with respect to the time $S_n\equiv S$ of 
eq.~(\ref{defPimoving}) and using the 
fact that $i\lambda_j$ $(j=1,\cdots, n)$ can be replaced $\partial_j$, we discover that $\Pi_{\eps} (\d;S)$ satisfies the Fokker-Planck (FP) equation
\be\label{FP35}
\frac{\partial \Pi_{\eps} (\d;S)}{\partial S}=\frac{1}{2}\frac{\partial^2\Pi_{\eps} (\d;S)}{\partial\delta^2},
\ee
(where we used the notation $\delta_n=\delta$). To determine the boundary condition to be imposed on the solution of \eq{FP35} we proceed as follows.
We  start from eq.~(\ref{defPimoving}), 
with $W$ given by eq.~(\ref{W}) and, shifting the variables $\delta_i$ $(i=1,\ldots, n)$ as
$\delta_i\to \delta_i-B(S_i)$, we obtain
\begin{eqnarray}
&&\Pi_{\eps} (\d_n+B_n;S_n)=
\int_{-\infty}^{0} d\d_1\ldots \int_{-\infty}^{0}d\d_{n-1}\nn\\
&&\times{1\over (2\pi\eps)^{n/2}}e^{ 
-{1\over 2\eps}\sum_{i=0}^{n-1} [\delta_{i+1}-\delta_i+B_n-B_{n-1}]^2
}\nn\\
&&= \int_{-\infty}^0 d\d_{n-1} {1\over \sqrt{2\pi\eps}} e^{-{1\over 2\eps}[\delta_n-\delta_{n-1}
+B_n-B_{n-1}]^2}\nn\\
&&\times\Pi_{\eps}(\delta_{n-1}+B_{n-1};S_{n-1})\, ,
\label{Pishifted}
\end{eqnarray}
where we used the notation $B_i\equiv B(S_i)$, so $B_n\equiv B_n$.
Now let $S_{n-1}=S$ so  $S_n=S+\eps$, and $ \delta_n+B(S)=\d, \delta_n-\delta_{n-1}=\Delta\d$.
For fixed $\d_n$, we have $d\d_{n-1}=-d(\Delta\d)$. By further taking the limit $\eps\to 0$ (assuming that $B(S)$ is a continuous and differentiable function), 
eq.~(\ref{Pishifted}) becomes
\be
\Pi_{\eps=0} (\d;S)=\int_{\delta-B(S)}^\infty d(\Delta\delta) \d_D(\Delta\d)\Pi_{\eps=0}(\d-\Delta\d; S)\,.
\ee
From this  relation we get the boundary condition. If $\delta=B(S)$ the integral is over half of the support of the Dirac delta and so $\Pi_{\epsilon=0}(B(S) ;S)=(1/2)\Pi_{\epsilon=0}(B(S) ;S)$
hence $\Pi_{\epsilon=0}(B(S) ;S)=0$.
Furthermore, if $\delta>B(S)$, the support of the Dirac delta is outside the integration limits and therefore
we conclude that
\be\label{Boundcond}
\Pi_{\epsilon=0}(\d ;S)=0 \qquad \textrm{for}\quad \delta\geq B(S)\,.
\ee
In the continuum limit the first-crossing rate is then given by
\begin{eqnarray}
{\cal F}(S)  &\hspace*{-2mm}=&\hspace*{-2mm} -\frac{\pa}{\pa S}
\int_{-\infty}^{B(S)}d\d\, \Pi_{\eps=0}(\d;S)\\
&\hspace*{-2mm}=&\hspace*{-2mm}-{dB(S)\over dS}\Pi_{\eps=0}(B(S),S)-\int_{-\infty}^{B(S)}d\d\, \frac{\partial\Pi_{\eps=0}(\d;S)}{\partial S}\, .\nonumber
\ees
The first term on the right-hand side vanishes because of the boundary condition, while the second term can be written in  a more convenient form using the FP equation (\ref{FP35}),
so
\bees
{\cal F}(S)&=&
-\frac{1}{2}\int_{-\infty}^{B(S)}d\d\, \frac{\partial^2\Pi_{\eps=0}(\d;S)}{\partial \d^2}\nn\\
&=&-\frac{1}{2}\left.\frac{\partial\Pi_{\eps=0}(\d;S)}{\partial \d}\right|_{\delta=B(S)}\, .
\label{firstcrossmoving}
\end{eqnarray}
To compute the probability
$\Pi_{\eps=0}(\delta_n,S_n)$ we proceed in the following way. At every $i$-th step of the path integral we Taylor expand the barrier around its final value
\be
B(S_i)=B(S_n)+\sum_{p=1}^\infty\frac{\Bnp}{p!}\, \left(S_i-S_n\right)^p\, ,
\ee
where
\be
\Bnp\equiv  \frac{d^p B(S_n)}{d S_n^p}\, , 
\ee
(so in particular $B_n^{(0)}=B(S_n)\equiv B_n$).
We now perform a shift in the variable $\delta_i$ ($i=1,\ldots,n-1$) in the path integral
\be
\label{shift}
\delta_i\rightarrow \delta_i-\sum_{p=1}^\infty\frac{\Bnp}{p!}
\, \left(S_i-S_n\right)^p\, ,
\ee
Then $\Pi_{\eps} (\d_n;S_n)$ can be written as 
\be
\Pi_{\eps} (\d_n;S_n)=
 \int_{-\infty}^{B_n} d\d_1\ldots \int_{-\infty}^{B_n}d\d_{n-1}\,\Dl \,\,e^Z
 \ee
 where
 \bees
 Z&=&i\sum_{i=1}^n\lambda_i\d_i
 -\frac{1}{2}\sum_{i,j=1}^n\lambda_{i}\lambda_{j}\,{\rm min}(S_i,S_j)\nn\\
&&
 +i\sum_{i=1}^{n-1}\lambda_i\sum_{p=1}^\infty\frac{\Bnp}{p!}
\, \left(S_i-S_n\right)^p .
\ees
We next expand 
\begin{eqnarray}
&&\hspace*{-5mm}{\rm exp}\left\{i\sum_{i=1}^{n-1}\lambda_i\sum_{p=1}^\infty\frac{\Bnp}{p!}
\, \left(S_i-S_n\right)^p\right\}\nn\\
&&\hspace*{-5mm}
\simeq 1+i\sum_{i=1}^{n-1}\lambda_i\sum_{p=1}^\infty\frac{\Bnp}{p!}\, \left(S_i-S_n\right)^p\\
&&\hspace*{-5mm}
-\frac{1}{2}\sum_{i,j=1}^{n-1}\lambda_i\lambda_j\sum_{p,q=1}^\infty\frac{\Bnp\Bnq}{p!q!}
\left(S_i-S_n\right)^p
\, \left(S_j-S_n\right)^q+\cdots\, ,\nn
\end{eqnarray}
and we  write $\Pi_{\eps} (\d_n;S_n)$ as 
\bees
\Pi_{\eps} (\d_n;S_n)&=&\Pi^{(0)}_{\eps} (\d_n;S_n)+\Pi^{(1)}_{\eps} (\d_n;S_n)\nn\\
&&+\Pi^{(2)}_{\eps} (\d_n;S_n)
+\cdots\, ,
\ees
where
\begin{eqnarray}
\Pi^{(0)}_{\eps=0} (\d_n;S_n)=\frac{1}{\sqrt{2\pi S_n}}\,
\[  e^{-\d_n^2/(2S_n)}- e^{-(2B_n-\d_n)^2/(2S_n)} \] ,
\end{eqnarray}
\bees
\Pi^{(1)}_{\eps} (\d_n;S_n)&=&\sum_{i=1}^{n-1}\int_{-\infty}^{B_n} d\d_1\ldots d\d_{n-1}
\sum_{p=1}^\infty\frac{\Bnp}{p!}
\label{Pi1}\\
&&\times\left(S_i-S_n\right)^p \partial_i W^{\rm gm}(\delta_0;\d_1,\ldots,\delta_n; S_n)\, ,\nn
\ees
and
\begin{eqnarray}
\label{s2}
&&\hspace*{-5mm}\Pi^{(2)}_{\eps} (\d_n;S_n)=\frac{1}{2}\sum_{i,j=1}^{n-1}\int_{-\infty}^{B_n} d\d_1\ldots d\d_{n-1}
\sum_{p,q=1}^\infty\frac{\Bnp\Bnq}{p!q!}\nn\\
&&\hspace*{-5mm}\times\left(S_i-S_n\right)^p
\left(S_j-S_n\right)^q
\partial_i\partial_j W^{\rm gm}(\delta_0;\d_1,\ldots,\delta_n; S_n)\, .
\end{eqnarray}
We have therefore formally expanded $\Pi_{\eps=0}(\d_n,S_n)$ in a series of terms
$\Pi^{(1)}_{\eps=0}$, $\Pi^{(2)}_{\eps=0}$, etc., in which each term is itself given by an infinite sum over indices $p, q,\ldots$. To proceed further, we must either perform some approximation, or identify a suitable small parameter, and organize 
the terms in a systematic expansion in such a small parameter. In the next  subsections we first discuss the approximation in which one can rederive the Sheth-Tormen result, and we will then compare it with two complementary, and more systematic, expansions.

\subsection{The Sheth-Tormen approximation}\label{sect:STbarr}

To attack the problem, a first idea is  to perform the integrals in \eqs{Pi1}{s2} approximating  $(S_n-S_i)^{p-1}\simeq S_n^{p-1}$ inside the integrals.  This is in fact equivalent to the approximation made  in \cite{lam}, see in particular their eq.~(20).
The detailed calculations, within our formalism,  are reported in
Appendix \ref{reproducingST} and one obtains the first-crossing rate for a
moving barrier 
\begin{eqnarray}
\label{total}
{\cal F}_{\rm ST}(S)&=&\frac{e^{-B^2(S)/(2S)}}{\sqrt{2\pi}S^{3/2}}
\sum_{p=0}^\infty\frac{(-S)^p}{p!} \frac{\partial^p B(S)}{\partial S^p}\, .
\end{eqnarray}
This expression agrees  with the one  suggested  in \cite{ST2002}. Notice that for the cases of constant barrier $B(S)=\delta_c$ and of a linear barrier $B(S)=\delta_c+\beta S$, which are the known examples where the first-crossing rate can be computed analytically
by solving exactly the FP equation in the presence of such a barrier (for the linear barrier see
\cite{Sheth1998} and  Section IX of \cite{Zentner}) the first-crossing rate (\ref{total}) reproduces the correct answer. When applied to the ellipsoidal barrier given in
eq.~(\ref{B(S)2}), and restricting the sum to $p\leq 5$, one recovers the ellipsoidal collapse result of 
\cite{ST2002}
\begin{eqnarray}
{\cal F}_{\rm ST}^{\rm{ell}}(S)&\simeq& 
\frac{\sqrt{a}\,\delta_c}{\sqrt{2\pi }S^{3/2}}e^{-B^2(S)/(2S)}\Bigg[1+ \nn\\
&&\left. +0.4\sum_{p=0}^5(-1)^p{0.6\choose p}
\left(\frac{S}{a \delta_c^2}\right)^{0.6}\right]\nn\\
&&\hspace*{-10mm}= \frac{\sqrt{a}\,\delta_c}{\sqrt{2\pi }S^{3/2}}e^{-B^2(S)/(2S)}\left[1+0.067\left(\frac{S}{a \delta_c^2}\right)^{0.6}\right].
\label{FST}
\end{eqnarray}
This procedure is,  however, not free from drawbacks. Indeed, the restriction of the sum to $p\leq 5$ is not justified and  is merely dictated by the fact that stopping arbitrarily 
the series at $p=5$ provides  a  a good fit to the N-body 
simulations.\footnote{We thank Ravi Sheth for discussions about this point.} However,    if the sum over $p$ is extended up to infinity the sum simply resums to $B(0)$ since, performing a Taylor expansion of $B(S_0-S)$ in powers of  $S$ and setting finally $S_0=S$, we have
\be
B(0)=\sum_{p=0}^\infty\frac{(-S)^p}{p!} \frac{\partial^p B(S)}{\partial S^p}\, .
\ee
Since $B(0)=\sqrt{a}\d_c$, we just end up with
\be
{\cal F}_{p=\infty}^{\rm{ell}}(S)=\frac{\sqrt{a}\,\delta_c}{\sqrt{2\pi }S^{3/2}}e^{-B^2(S)/(2S)}\, ,
\ee
so the correction term $\sim S^{0.6}$ in \eq{FST} seems an artifact of stopping the sum to $p=5$.
This is a rather puzzling result, since this correction is known to fit well the data, and is widely used in the literature. This
calls for a different and more rigorous approach where the
integrals are performed without the approximation $(S_n-S_i)^{p-1}\simeq S_n^{p-1}$. We discuss two different possible  approaches in the next two subsection.

\subsection{Expansion of $\Pi_{\eps}(\d,S)$ in derivatives of $B(S)$}
\label{sect:der}

In order to develop a more systematic expansion, we first
consider the  case of a barrier $B(S)$ which is slowly varying with $S$. In this case, the small parameters are the derivatives of the function $B(S)$.

At first one might think that such an approximation, altough useful in some cases, would not apply to  the barrier which corresponds to the the ellipsoidal collapse, \eq{B(S)2}. In this case infact $B_{\rm el}(S)$ is given by a constant plus a term proportional to $S^{\g}$ with $\g\simeq 0.6<1$, and therefore already its first derivative, which is proportional to $S^{\g-1}$ is large at sufficiently small $S$, and formally even diverges as $S\ra 0$. However one should not forget that, in practice, even the largest galaxy clusters than one finds in observations, as well as in large-scale $N$-body simulations, have typical masses smaller than about $10^{15} h^{-1}\msun$ which, in the standard $\Lambda$CDM
cosmology, corresponds to values of $S=\s^2(M)\gsim 0.35$, see e.g. Fig.~1 of \cite{Zentner}.  Even for such a value, which is the smallest in which we are interested, the value of $B'_{\rm el}(S)$ is just of order 0.3 which means that, in the range of masses of interest, the barrier of ellipsoidal collapse can be considered as slowly varying.

We therefore expand $\Pi_{\eps}(\d_n;S_n)$ in powers of the derivatives of the barrier, keeping terms with the same number of derivatives, so for instance a term proportional to $d^2B/dS^2$ is taken to be of the same order as $(dB/dS)^2$. 
Working up to terms of second order in the derivatives we get 
\bees
\Pi_{\eps}(\d_n;S_n) &=& \Pi_{\eps}^{(0)}(\d_n;S_n)+ \Pi^{(a)}_{\eps}(\d_n;S_n) + 
\Pi^{(b)}_{\eps}(\d_n;S_n) \nn\\
&&+\Pi^{(c)}_{\eps}(\d_n;S_n) \, ,
\ees
where
\bees
\Pi^{(a)}_{\eps}(\d_n;S_n)&=&\sum_{i=1}^{n-1}B_n'(S_i-S_n)\\
&&\times\int_{-\infty}^{B_n}d\d_1\ldots d\d_{n-1}
\pa_i W^{\rm gm}\, ,\nn\\
\Pi^{(b)}_{\eps}(\d_n;S_n)&=&\frac{1}{2}\sum_{i=1}^{n-1}
B_n'' (S_i-S_n)^2\\
&&\times \int_{-\infty}^{B_n}d\d_1\ldots d\d_{n-1}
\pa_i W^{\rm gm}\, ,\nn\\
\Pi^{(c)}_{\eps}(\d_n;S_n)&=&\frac{1}{2}\sum_{i,j=1}^{n-1}\(B_n'\)^2 (S_i-S_n)(S_j-S_n)\\
&&\times\int_{-\infty}^{B_n}d\d_1\ldots d\d_{n-1}
\pa_i\pa_j W^{\rm gm} ,\nn
\ees
and we used a prime to denote the derivatives 
of $B(S_n)$ with respect to $S_n$.
Observe that $\Pi^{(a)}$ and  $\Pi^{(b)}$ are linear in the first and second
derivative, respectively, and come from the terms $p=1,2$ of
$\Pi^{(1)}$, while $\Pi^{(c)}$ is quadratic in the first derivative, and is the term $p=q=1$ of
$\Pi^{(2)}$.

In Appendix~\ref{App:A} we compute these three terms, in the continuum limit, using the techniques developed in MR1. For the first term we find
\be
\Pi^{(a)}_{\eps=0}(\d_n;S_n)=
-2 B_n'\, \frac{(B_n-\d_n)}{\sqrt{2\pi S_n}} e^{-(2B_n-\d_n)^2/(2S_n)}\, .
\ee
Observe that it satisfies the boundary condition $ \Pi^{(a)}_{\eps=0}(\d_n;S_n)=0$ when $\d_n=B_n$, as it should. For the 
second term we get
\bees
&&\hspace*{-4mm}
\Pi^{(b)}_{\eps=0}(\d_n;S_n)= \frac{1}{2\pi} B_n'' (B_n-\d_n)\, \\
&&\hspace*{-4mm}\times\[ \sqrt{2\pi S_n} e^{-(2B_n-\d_n)^2/(2S_n)}
-\pi B_n {\rm Erfc}\(\frac{2B_n-\d_n}{\sqrt{2S_n}}\)\]\, ,\nn
\ees
and again vanishes linearly as $\d_n\ra B_n$.
The third term is given by
\be
\Pi^{(c)}_{\eps=0}(\d_n;S_n)=
-2
\(B_n'\)^2\,\frac{(B_n-\d_n)^2}{\sqrt{2\pi S_n}} e^{ -(2B_n-\d_n)^2/(2S_n)}\hspace{-1mm} ,
\ee
and vanishes quadratically as $\d_n\ra B_n$. This means that in the end it does not contribute to the first-crossing rate, since, using \eq{firstcrossmoving}, the latter is given by the derivative of 
$\Pi_{\eps=0} (\d_n;S_n) $ with respect to $\d_n$, evaluated in $\d_n=B_n$.

It is interesting to check explicitly that this solution for $\Pi (\d_n;S_n)$ satisfies the FP equation, up to order to which we have computed, i.e. up to terms of second order in the derivatives of the barrier, included.
Define the FP operator
\be
\hat{D}=\frac{\pa}{\pa S_n}-\frac{1}{2}\frac{\pa^2}{\pa\d_n^2}\, ,
\ee
and define $f^{(0)}, \ldots f^{(c)}$ from
\be
\hat{D}\Pi_{\eps=0}^{A}(\d_n;S_n) =
\sqrt{\frac{2}{\pi}}\, \frac{1}{S_n^{3/2}} e^{-(2B_n-\d)^2/(2S_n)}
 f^{A}\, ,
\ee
where $A=(0),(a),(b),(c)$ so, up to terms of second order (included)  in the derivatives of the barrier,
\bees
\hat{D}\Pi_{\eps=0}(\d_n;S_n) &=&
\sqrt{\frac{2}{\pi}}
\, \frac{1}{S_n^{3/2}} e^{-(2B_n-\d_n)^2/(2S_n)}\nn\\
&&\times [f^{(0)}+f^{(a)}+f^{(b)}+f^{(c)}]
\, .
\ees
Inserting the expressions for  $\Pi_{\eps=0}^{(0)}, \Pi_{\eps=0}^{(a)},\Pi_{\eps=0}^{(b)},\Pi_{\eps=0}^{(c)}  $ computed above we get
\bees
f^{(0)}&=&(2B_n-\d_n) B_n'\, ,\\
f^{(a)}&=&-(2B_n-\d_n) B_n' -S_n (B_n-\d_n) B_n'' \nn\\
&&+[2(B_n-\d_n)(2B_n-\d_n)-S_n] (B_n')^2\\
f^{(b)}&=& S_n (B_n-\d_n) B_n'' +{\cal O}(B_n''',B_n'B_n'', (B_n')^3)\\
f^{(c)}&=&-[2(B_n-\d_n)(2B_n-\d_n)-S_n] (B_n')^2\nn\\
&&+{\cal O}(B_n''',B_n'B_n'', (B_n')^3)
\ees
Therefore the sum $\Pi^{(0)}_{\eps=0}+\Pi^{(a)}_{\eps=0}+\Pi^{(b)}_{\eps=0}+
\Pi^{(c)}_{\eps=0}$ satisfies the FP equation, modulo terms of third order in the derivative of the barrier. 

\begin{figure}
\centering
\includegraphics[width=0.45\textwidth]{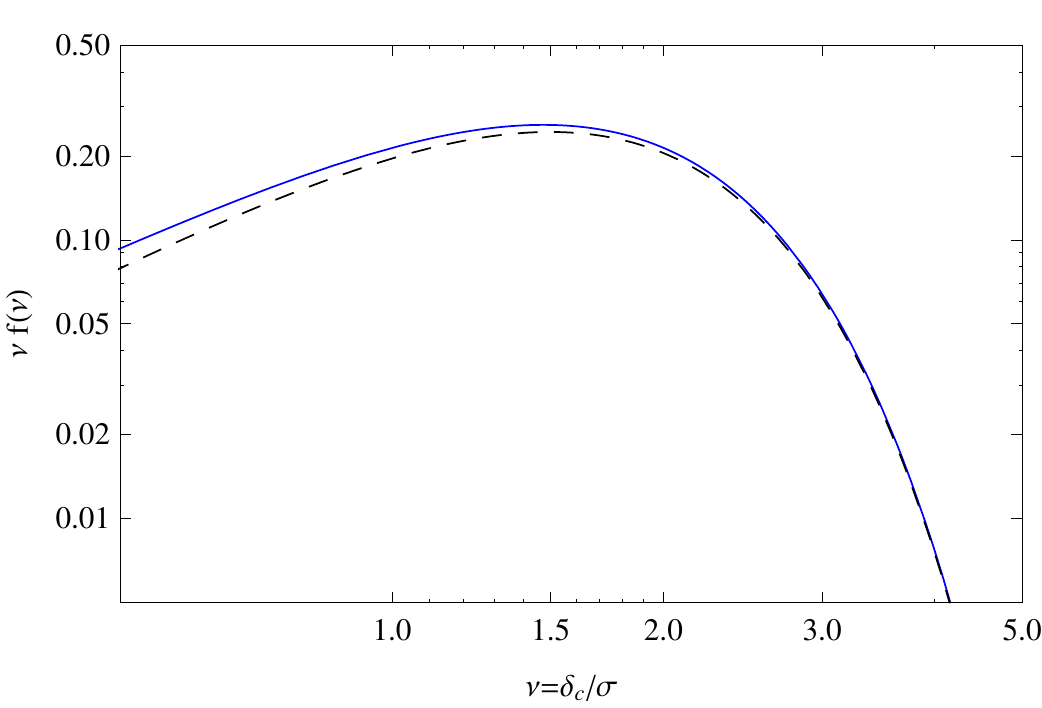}
\caption{
\label{figure:Fderivatives}
The Sheth-Tormen first-crossing rate for the ellipsoidal  barrier 
$\mathcal{F}_{\rm ST}^{\rm ell}$ (dashed black line), compared to the first-crossing rate
${\cal F}^{(2)}_{\rm der}$ (solid blue line)
obtained from the expansion in derivatives of the barrier, as a function of $\nu$.
}
\end{figure}

\begin{figure}
\centering
\includegraphics[width=0.45\textwidth]{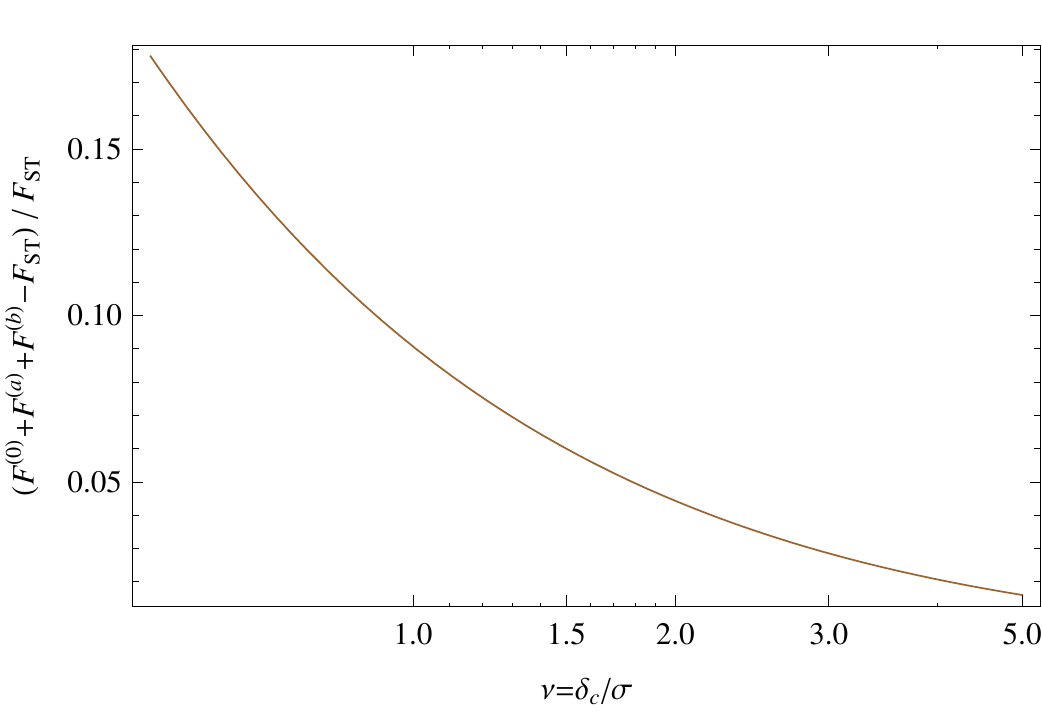}
\caption{
\label{figure:ratioF}
The ratio
$({\cal F}_{\rm der}^{(2)}-\mathcal{F}_{\rm ST})/\mathcal{F}_{\rm ST}$, as a function of $\nu$.
}
\end{figure}

 The first-crossing rate is then readily evaluated through eq.~(\ref{firstcrossmoving}). 
The zero-th order contribution from $\Pi^{(0)}_{\eps=0}$ is 
\be
\label{F0}
{\cal F}^{(0)}(S)=\frac{B(S)}{\sqrt{2\pi}S^{3/2}}e^{-B^2(S)/(2S)}\, ,
\ee
while the higher orders give
\begin{eqnarray}
\mathcal{F}^{(a)}(S)&=&-{B'(S)\over\sqrt{2\pi S}}e^{-B(S)^2/(2S)}\,,\label{Fa}\\
\mathcal{F}^{(b)}(S)&=&{B''(S)\over{4\pi}}\\
&&\hspace*{-5mm}\times \left\{\sqrt{2\pi S}e^{-B(S)^2/(2S)}\
-\pi B(S)\textrm{Erfc}\left[{B(S)\over 2S}\right]\right\}\label{Fb}
\,.\nn
\end{eqnarray}
and  $\mathcal{F}^{(c)}=0$, as already mentioned.
In Fig.~\ref{figure:Fderivatives} we compare the Sheth-Tormen first crossing rate
$\mathcal{F}_{\rm ST}(S)$ to the quantity 
\be
{\cal F}^{(2)}_{\rm der}(S)={\cal F}^{(0)}(S)+\mathcal{F}^{(a)}(S)+\mathcal{F}^{(b)}(S)\, ,
\label{Fdertotal}
\ee
i.e. to  the first crossing rate obtained by performing the expansion in derivatives of the barrier, up to second order (included) in the derivatives, while in 
Fig.~\ref{figure:ratioF} we plot the relative difference
$({\cal F}^{(2)}_{\rm der}-\mathcal{F}_{\rm ST})/\mathcal{F}_{\rm ST}$. We see that the two results agree perfectly at large values of $\nu$ (i.e. at large masses), and they still agree to better than 10\% down to $\nu=1$. 

The fact that the ${\cal F}^{(2)}_{\rm der}$ is numerically quite close to 
$\mathcal{F}_{\rm ST}$
provides a more satisfying derivation of  the ST mass function, showing that the 
approximation $(S_n-S_i)^{p-1}\simeq S_n^{p-1}$, together with
the truncation  to $p=5$  of the series in \eq{total}, in the end gives a simple analytic formula which is numerically quite close to the result of a derivation based on a systematic expansion.

\begin{figure}
\centering
\includegraphics[width=0.45\textwidth]{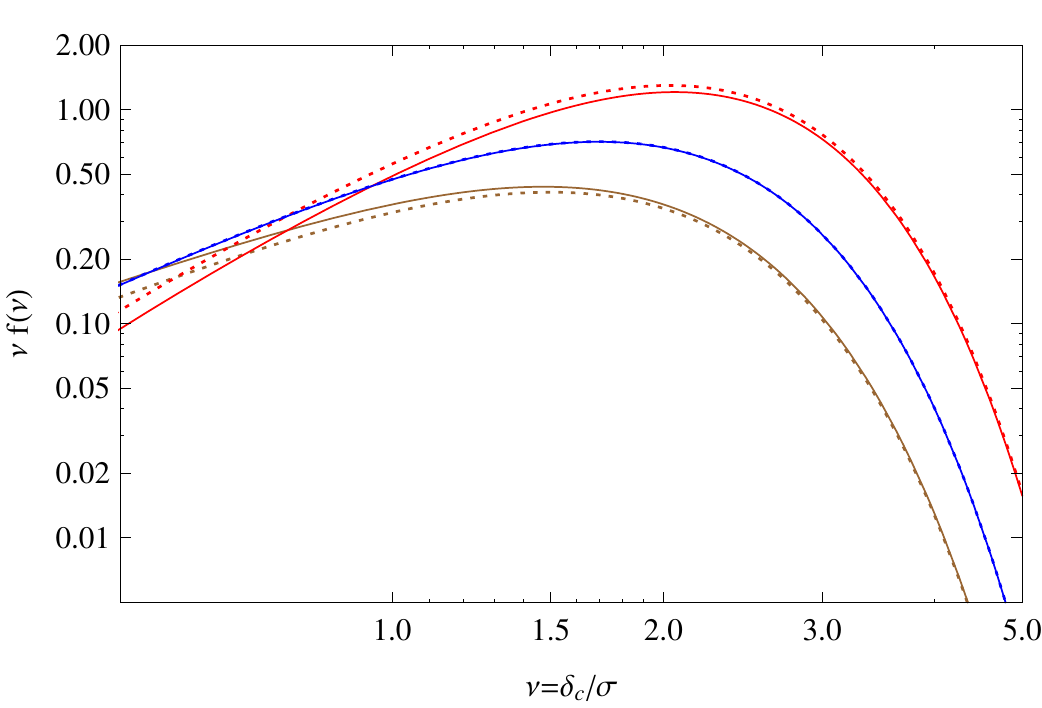}
\caption{
\label{3barriers}
First-crossing rate for
filaments (blue), sheets (red) and halos (brown). The dotted lines refer to the ST approximation  (\ref{total}) with $p\leq 5$, while
the continuous ones refer to our result  (\ref{Fdertotal}).}
\end{figure}

For comparison, we also report in Figure \ref{3barriers} the first-crossing rate for
filaments (blue), sheets (red) and halos (brown). The dashed lines refer to the ST approximation (\ref{total}) with $p\leq 5$, while
the continuous ones refer to our result (\ref{Fdertotal}). 

\subsection{Expansion of $\Pi_{\eps=0}(\d_n,S_n)$ in powers of $(B_n-\d_n)$}
\label{subsection:expansion}

In this subsection we describe a different expansion scheme, which allows us to resum a large number of terms. The basic idea is that,
even if
the computation of the distribution function $\Pi$  can be interesting  by itself in a more general context (since the probability distribution of a random walk in the presence of a moving barrier is a problem interesting in its own right in statistical physics), for the computation of the halo mass function we are really interested only in the first-crossing rate. Then \eq{firstcrossmoving} shows that, in the Gaussian and Markovian case, we only need the derivative  $\pa\Pi/\pa\d_n$ evaluated at
$\d_n=B_n$. As shown in \eq{Boundcond},
$\Pi(\d_n,S_n)$ vanishes at $\d_n=B_n$, so its Taylor expansion around $\d_n=B_n$ 
starts from a term linear in  $(\d_n-B)$, followed by terms of order $(\d_n-B)^2$, etc. When we compute $\pa\Pi/\pa\d_n$ in
$\d_n=B_n$, the terms quadratic and higher-order in $(\d_n-B)$, give zero, so we do not need the full function $\Pi$, but only the term linear in $(\d_n-B)$ in its Taylor expansion around $\d_n=B_n$. This simplifies our task considerably.  

\begin{figure}
\centering
\includegraphics[width=0.45\textwidth]{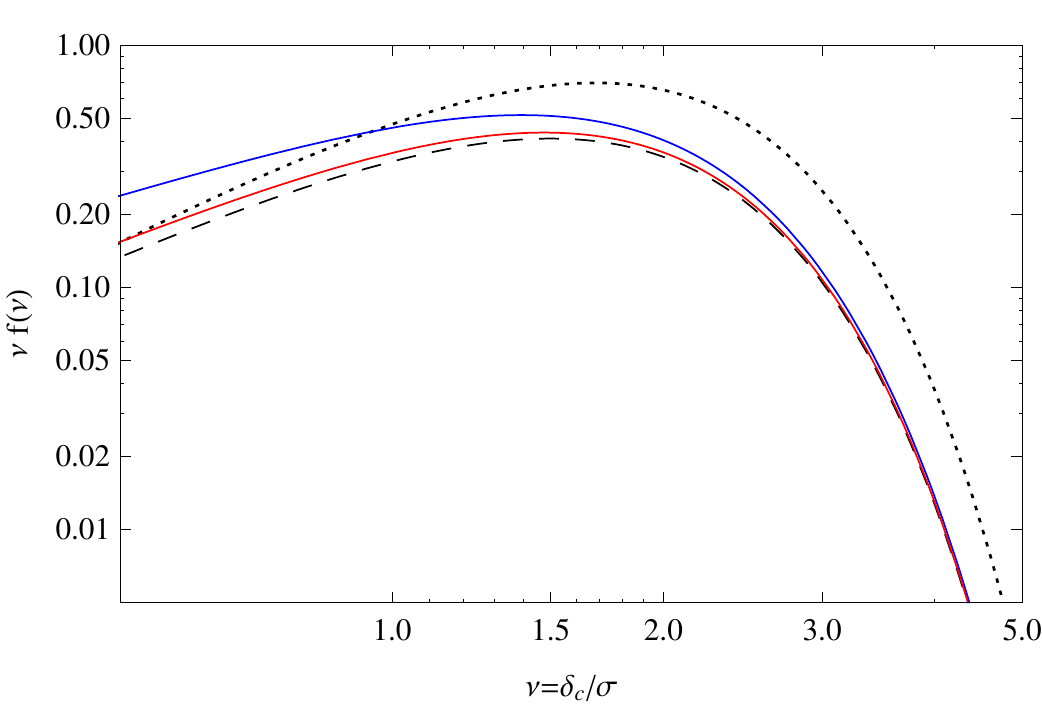}
\caption{
\label{figure:compareF}
First-crossing rate for the ellipsoidal  barrier  (\ref{B(S)2}).
$\mathcal{F}_{\rm ST}$ (dashed black), $\mathcal{F}^{(0)}$ (solid blue),
$\mathcal{F}^{(0)}+\mathcal{F}^{(1)}+{\cal F}^{(2)}+\cdots$ (solid red). The spherical collapse model, with the
same value of $a=0.707$, corresponds to the dotted black line.}
\end{figure}

We  first compute the part linear in $(\d_n-B_n)$ of
$\Pi^{(1)}$. Using the results of the previous section, in particular eqs. (\ref{byparts}), 
(\ref{Pigammafinal}) and (\ref{Pigammafinalbis}), 
  $\Pi^{(1)}_{\eps} (\d_n;S_n)$ can be rewritten as
\begin{eqnarray}
\label{a}
\Pi^{(1)}_{\eps=0} (\d_n;S_n)&=&\frac{B_n(B_n-\delta_n)}{\pi}\sum_{p=1}^\infty\frac{(-1)^p}{p!}\Bnp
\\
&&\times\int_0^{S_n}d S_i\frac{\left(S_n-S_i\right)^{p-(3/2)}
}{S_i^{3/2}}\nn\\
&&\times e^{-B_n^2/(2S_i)}e^{-(B_n-\delta_n)^2/[2 (S_n-S_i)]}\, .\nn
\end{eqnarray}
For $p=0,1$ this integral can be computed analytically, see appendix~\ref{App:B}, but for $p\geq 2$ we have not been able to compute it exactly. However, for our purposes it is sufficient to observe that in
this expression for $\Pi^{(1)}_{\eps=0}$ there is  already  a factor $(B_n-\d_n)$ in front of the 
integral over $dS_i$, and   the integral converges at $S_i=S_n$ for all $p\geq 1$, even if in the integrand we set $\d_n=B_n$. Therefore
\begin{eqnarray}\label{aaa2}
\Pi^{(1)}_{\eps=0} (\d_n;S_n)&=&\frac{B_n(B_n-\delta_n)}{\pi}\sum_{p=1}^\infty\frac{(-1)^p}{p!}\Bnp\\
&&\times\int_0^{S_n}d S_i\frac{\left(S_n-S_i\right)^{p-(3/2)}
}{S_i^{3/2}} e^{-B_n^2/(2S_i)}\nn\\
&&+{\cal O} (B_n-\delta_n)^2\, .\nn
\end{eqnarray}
In appendix~\ref{App:B} we show that for $p=1$  this integral is elementary while for $p\geq 2$ it can be computed in terms of  the confluent hypergeometric function $U(a,b,z)$. As a result,
\bees
&&\hspace*{-5mm}
\Pi^{(1)}_{\eps=0} (\d_n;S_n)={\sqrt{2}\over \pi}\, \frac{B_n-\d_n}{S_n^{1/2}}\,
e^{-B_n^2/(2S_n)}\nn \\
&&\hspace*{-3mm}\times\[\sum_{p=1}^{\infty}\frac{(-1)^p}{p!}\Bnp
S_n^{p-1}\, \Gamma\(p-\frac{1}{2}\)  U\(p-1,\frac{1}{2},\frac{B_n^2}{2S_n}\)
\]\nn\\
&&\hspace*{-5mm}+{\cal O} (B_n-\delta_n)^2\, ,
\ees
where the term $p=1$ can be written in a more elementary form using
$U(0,b,z)=1$ and $\Gamma(1/2)=\sqrt{\pi}$.
{Along the same lines, we have also computed the generic $m$-th order ($m\geq 1$) of the expansion of $\Pi_{\eps=0}$ 
(see App.~\ref{App:C}), at the linear order in $B_n-\delta_n$, and it is given by
\begin{eqnarray}
\Pi_{\epsilon=0}^{(m)}&=&{(B_n-\delta_n)e^{-{B_n^2\over 2S_n}}\over m! \,2^{{m\over 2}-1}\pi^{3-m\over 2}}
\sum_{p_1,\ldots,p_m=1}^\infty (-1)^{\sum_{k=1}^m p_k+m+1}\nn\\
&\times&{B_n^{(p_1)}\cdots B_n^{(p_m)}
\over p_1!\cdots p_m!} c_{p_2,\ldots, p_m}S_n^{\sum_{k=1}^m p_k-{m\over 2}-1}\nn\\ 
&\times& \Gamma\left(\sum_{k=1}^m p_k-{m\over 2}\right)
U\left(\sum_{k=1}^m p_k-{m+1\over 2},{1\over 2}, {B_n^2\over 2S_n}\right)\nn\\
&+&\mathcal{O}(B_n-\delta_n)^2,
\label{Pimresult}
\end{eqnarray}
where the coefficients $c_{p,q,\cdots}$ can be computed by the recursion relations (\ref{cpm})-(\ref{cp2}).
This expression is useful for numerical evaluation, but  not very illuminating from an analytic point of view. 
So it can be useful to keep in mind that in the limit 
$2S_n\ll B_n^2$, i.e. for large halo masses, 
the confluent hypergeometric $U$ function simplifies to
\be
U\(k,\frac{1}{2},\frac{B_n^2}{2S_n}\)\simeq \(\frac{2S_n}{B_n^2}\)^{k}
\[1+{\cal O}\(\frac{2S_n}{B_n^2}\)\] \, .
\ee
The total probability is  given by $\Pi=\sum_{m=0}^\infty \Pi^{(m)}$.
We have not been able to resum all the terms of the expansion, but
the first few terms are sufficient for the first-crossing rate.
In fact,  the first-crossing rate is readily evaluated  through eq.~(\ref{firstcrossmoving}). The zero-th order contribution from $\Pi^{(0)}_{\eps=0}$ is given by eq.~(\ref{F0})
while higher-order contributions $\mathcal{F}^{(m)}$ are obtained  from 
$\Pi_{\epsilon=0}^{(m)}$ in eq.~(\ref{Pimresult}), and are easily evaluated numerically.
In Fig.~\ref{figure:compareF}, we plot ${\cal F}^{(0)}$ (blue) and ${\cal F}^{(0)}+{\cal F}^{(1)}+{\cal F}^{(1)}
+\cdots$
(red), for the ellipsoidal  barrier  given in eq.~(\ref{B(S)2}). 
 We deduce that the sum for $\Pi$ converges quickly and
the  terms after the second one contribute negligibly to the first-crossing rate.
It is therefore an excellent approximation to consider the first-crossing rate for a generic moving barrier $B(S)$
as given by ${\cal F}^{(0)}+{\cal F}^{(1)}$, i.e.
\begin{eqnarray}
\label{Ftotal}
{\cal F}(S)&=&\frac{e^{-B^2(S)/(2S)}}{\sqrt{2\pi}S^{3/2}}\bigg[
B(S)\nn\\
&&\hspace*{-20mm}
\left.+\sum_{p=1}^\infty\frac{(-S)^p}{p!} \frac{\partial^p B(S)}{\partial S^p}
{\Gamma\(p-\frac{1}{2}\)\over \sqrt{\pi}}  U\(p-1,\frac{1}{2},\frac{B_n^2}{2S_n}\)
\right]\, .
\end{eqnarray}
For comparison, we  also report in 
Fig.~ \ref{figure:compareF} the first-crossing rate of the spherical collapse model (dotted line) and the \cite{ST2002} result of eq.~(\ref{FST}) (dashed line). 
Note also that eq.~(\ref{Ftotal}) reproduces the exact known results for the cases of constant and linear barrier shapes.
}
{
It is also interesting to note that ${\cal F}(S)$ in eq. (\ref{Ftotal})  and the  rate 
${\cal F}^{(2)}_{\rm der}(S)$ computed in the previous section
differ by less than 5\% for $\nu\geq 0.2$, for the ellipsoidal barrier (\ref{B(S)2}). It is then reassuring to see that 
our two approaches to  the computation of the first-crossing rate lead to consistent results, and their difference allows us to get a quantitative idea of the theoretical error in the computation. The fact that both results are numerically quite close to the ST mass function also provides a more satisfying justification  of  the ST mass function itself.
}

Armed with these results, we may now proceed to evaluate the halo mass function in the case in which
non-Gaussianity (NG) is present.

\section{The ellipsoidal collapse and non-Gaussianity}\label{sect:ellNG}

Deviations from Gaussianity are  encoded, e.g., in the connected 
 three- and four-point correlation functions which are dubbed the bispectrum
and the trispectrum,
respectively. A phenomenological way of parametrizing the level of NG is to
expand the fully 
 non-linear primordial Bardeen
 gravitational potential $\Phi$ in powers of the linear gravitational potential
$\Phi_{\rm L}$
 \be
 \label{phi}
 \Phi=\Phi_{\rm L}+f_{\rm NL}\left(\Phi_{\rm L}^2-\langle\Phi_{\rm
L}^2\rangle\right)\, .
 \ee
The  dimensionless
quantity  $f_{\rm NL}$  sets the
magnitude of the three-point 
correlation function  (\cite{bartoloreview}). If the process generating the primordial
NG  
is local in space, the parameter  $f_{\rm NL}$ in Fourier
space is
independent of the momenta
entering the corresponding  correlation functions; if instead the process which
generates the
primordial cosmological perturbations is
non-local in space, like in 
models of inflation with non-canonical kinetic terms, $f_{\rm NL}$ 
acquires a dependence on the momenta. 
The strongest current
limits on the strength of local NG set the $f_{\rm NL}$
parameter to be in the range $-4<f_{\rm NL}<80$ at 95\% confidence level
\citep{zal}. 

In MR3 the effect of primordial NG on the 
halo mass function  was computed, using 
excursion set theory, for the case of a spherical collapse with constant barrier. In
the presence of NG  the stochastic evolution of the smoothed density field, as a function of  the smoothing scale,  is
non-Markovian and beside ``local'' terms that generalize Press-Schechter (PS) theory,
there are also   ``memory'' terms, whose effect on the mass function have been   computed using the  formalism developed in MR1.
When computing the effect of the three-point correlator on the mass function, a PS-like approach which consists in neglecting the cloud-in-cloud problem and in multiplying the final result by a fudge factor $\simeq 2$, is in principle not justified. Indeed, when computed correctly in the framework of excursion set theory, the ``local" contribution vanishes (for all odd-point  correlators
the contribution of the image Gaussian cancels the Press-Schechter contribution rather than adding up), and the result
comes entirely from  non-trivial memory terms which are absent in PS theory. However it turns out that, in the limit of large halo masses, where the effect of non-Gaussianity is more relevant,
these memory terms give a contribution  which is the the same as that computed naively with PS theory, plus subleading terms depending on derivatives of the three-point correlator. 

The goal of this section is to compute, using excursion set theory, the halo mass function in the presence of NG and for the ellipsoidal collapse, thus extending the  findings of MR3 obtained for the spherical collapse.  This computation is motivated by the fact that in the literature the halo mass
function for the more realistic case of the ellipsoidal collapse is obtained, when NG is present, by multiplying the first-crossing rate  (\ref{FST}) by a form factor ${\cal R}(f_{\rm NL},S)$ obtained
by dividing the first-crossing rates with and without NG for the PS spherical collapse case
\begin{eqnarray}
{\cal F}_{\rm ST}(f_{\rm NL},S)&=&{\cal F}_{\rm ST}(f_{\rm NL}=0,S)
{\cal R}(f_{\rm NL},S)\nn\\
&=&{\cal F}_{\rm ST}(f_{\rm NL}=0,S)\frac{{\cal F}_{\rm PS}(f_{\rm NL},S)}{{\cal F}_{\rm PS}(f_{\rm NL}=0,S)}\, .
\label{FSTNG}
\end{eqnarray}
This procedure has however no rigourous justification and its validity should be tested with an explicit
computation.

Similarly to the Gaussian  case, the probability of arriving in $\d_n$ in a
``time'' $S_n$, starting from the initial value $\delta_0=0$, without ever
going above the threshold, in the presence of NG is given by 
\begin{eqnarray}
\Pi_{\eps} (\d_n;S_n)
& \equiv&\int_{-\infty}^{B(S_1)} d\d_1\ldots \int_{-\infty}^{B(S_{n-1})}d\d_{n-1}\, \nn\\
&&\times W_{\rm NG}(\d_0;\d_1,\ldots ,\d_{n-1},\d_n;S_n).
\label{NG}
\end{eqnarray}
where 
\begin{eqnarray}
W_{\rm NG} (\d_0;\d_1,\ldots ,\d_n;S_n)=\int{\cal D}\lambda\nn\\
\times\exp\left\{ i\sum_{i=1}^n\lambda_i\d_i-\frac{1}{2}\sum_{i,j=1}^n
\lambda_{i}\lambda_{j}\,
{\rm min}(S_i,S_j)\right\}\nn\\
\times\exp\left\{\frac{(-i)^3}{6}\sum_{i,j,k=1}^{n}\langle\d_i\d_j\d_k\rangle_c\lambda_i\lambda_j\lambda_k
\right\}\, .
\end{eqnarray}
We now perform the shift (\ref{shift}) in the $\delta_i$ $(i=1,\cdots,n-1)$ variables and expand the
NG contribution to first order
\begin{eqnarray}
\Pi_{\eps} (\d_n;S_n)&=&
\Pi^{(0)}_{\eps=0} (\d_n;S_n)+\Pi^{(1)}_{\eps=0} (\d_n;S_n)\nn\\
&+&\Pi^{(2)}_{\eps=0} (\d_n;S_n)+\cdots 
\nonumber\\
&-&\frac{1}{6}
\int_{-\infty}^{B_n} d\d_1\ldots \int_{-\infty}^{B_n}d\d_{n-1}\, \sum_{i,j,k=1}^{n}\nn\\
&& \langle\d_i\d_j\d_k\rangle_c\partial_i\partial_j\partial_k
W_{\rm mb}(\d_0;\d_1,\ldots ,\d_{n-1},\d_n;S_n) , \nonumber\\
\label{defPimovingNG}
\end{eqnarray}
where $W_{\rm mb}$ is the probability density in the space of trajectories  with a moving barrier, so that
\begin{eqnarray}
&&\hspace*{-5mm}\int_{-\infty}^{B_n} d\d_1\ldots \int_{-\infty}^{B_n}d\d_{n-1}\, 
W_{\rm mb}(\d_0;\d_1,\ldots ,\d_{n-1},\d_n;S_n)\nn\\
&&\hspace*{-5mm}=\Pi^{(0)}_{\eps=0} +\Pi^{(1)}_{\eps=0} +\Pi^{(2)}_{\eps=0} +\cdots\, .
\end{eqnarray}
In principle the contribution from NG   can be computed separating the various contributions to the sum
according to whether an index is equal or smaller than $n$. In this
way, however, the computations faces some technical
difficulties.
Fortunately, as discussed in MR3, the problem simplifies considerably in the limit of large
halo masses, which is just the physically interesting limit. Large
masses mean small values of $S_n$. The arguments
$S_i, S_j$ and
$S_k$ in the correlator
$\langle\d_i\d_j\d_k\rangle \equiv
\langle\d(S_i)\d(S_j)\d(S_k)\rangle_c$
range over the interval $[0,S_n]$ and, if $S_n$ goes to zero, we
can expand the correlator in a multiple Taylor series
around the point $S_i=S_j=S_k=S_n$.
We introduce the notation
\begin{eqnarray}
&G_3^{(p,q,r)}(S_n)\equiv&\nn\\
&\[\frac{d^p}{dS_i^p}\frac{d^q}{dS_j^q}\frac{d^r}{dS_k^r}
\langle\d(S_i)\d(S_j)\d(S_k)\rangle_c\]_{S_i=S_j=S_k=S_n}.&
\end{eqnarray}
Then
\bees
\label{pqrs}
\langle\d(S_i)\d(S_j)\d(S_k)\rangle
&=&
\sum_{p,q,r=0}^{\infty} \frac{(-1)^{p+q+r}}{p!q!r!}(S_n-S_i)^{p}\\
&\times&
(S_n-S_j)^{q}(S_n-S_k)^{r}G_3^{(p,q,r,s)}(S_n)\, .\nn
\ees
The leading contribution to the halo mass function 
is given by the term in \eq{pqrs} with $p=q=r=0$ and we neglect subleading contributions, which can be computed with the same technique developed in MR3. The discrete sum reduces to
 $\langle\delta_n^3\rangle_c \sum_{i,j,k=1}^{n}\partial_i\partial_j\partial_k$ and
we can split it as 
\begin{eqnarray}
\sum_{i,j,k=1}^{n}\partial_i\partial_j\partial_k&=&\partial_n^3+
3\sum_{i,j=1}^{n-1}\partial_i\partial_j\partial_n+3\sum_{i=1}^{n-1}\partial_i\partial_n^2\nn\\
&&+
\sum_{i,j,k=1}^{n-1}\partial_i\partial_j\partial_k\, .
\label{sumijk}
\end{eqnarray}
When applying these derivatives to the $W_{\rm mb}$, one can 
 use the  identities  proven in MR1 and MR3, namely
\begin{eqnarray}
\sum_{i=1}^{n-1}
\int_{-\infty}^{B_n} d\d_1\ldots d\d_{n-1}\, 
\pa_iW_{\rm mb}=\frac{\pa}{\pa B_n}
\Pi_{\eps=0}\, ,
\label{pa1xc}
\end{eqnarray}
\begin{eqnarray}
\sum_{i,j=1}^{n-1}
\int_{-\infty}^{B_n} d\d_1\ldots d\d_{n-1}\, 
\pa_i\pa_jW_{\rm mb}=\frac{\pa^2}{\pa B^2(S_n)}
\Pi_{\eps=0}\, ,
\label{pa2xc}
\end{eqnarray}
and 
\begin{eqnarray}
\sum_{i,j,k=1}^{n-1}
\int_{-\infty}^{B_n} d\d_1\ldots d\d_{n-1}\, 
\pa_i\pa_j\pa_kW_{\rm mb}=
\frac{\pa^3}{\pa B^3(S_n)}
\Pi_{\eps=0}\, .
\label{pa4xc}
\end{eqnarray}
The probability density (\ref{defPimovingNG}) calculated in this way vanishes at the
barrier point $\delta_n=B_n$, when one properly expands the $\Pi_{\eps=0}$ according
to one of the two methods described in the previous sections. 
This is a good check of the procedure we adopted and
is necessary when evaluating the first-crossing rate.

The calculation of the first-crossing rate proceeds by integrating the probability density
over $\delta_n$ and then taking the derivative with respect to $S_n$. This is fortunate because we can directly compute
\begin{eqnarray}
&&\sum_{i,j,k=1}^{n}
\int_{-\infty}^{B_n} d\d_1\ldots d\d_{n}\,
\pa_i\pa_j\pa_k W_{\rm mb}\nn\\
&&=\frac{\pa^3}{\pa B^3(S_n)}
\int_{-\infty}^{B_n} d\d_n \Pi_{\eps=0}
\end{eqnarray}
We choose two different expansions for $\Pi$. The expansion in derivatives of Sect.~\ref{sect:der}
gives
\begin{eqnarray}
&&\hspace*{-5mm}\frac{\pa^3}{\pa B^3(S_n)}
\int_{-\infty}^{B_n} d\d_n\,
\left(\Pi^{(0)}_{\eps=0} +\Pi^{(a)}_{\eps=0} +\Pi^{(b)}_{\eps=0}+\Pi^{(c)}_{\eps=0} +\cdots\right)\nonumber\\
&&\hspace*{-5mm}=
\frac{2}{\sqrt{2\pi}S^{5/2}} e^{-\frac{B^2}{2 S}} \left[-S+B^2+S B' \left(B+2 S B'\right)\right]
\nn\\
&&
-\frac{3}{2} \textrm{Erfc}\left[\frac{B}{\sqrt{2S}}\right] B''\,,
\label{pa3xc}
\end{eqnarray}
while the expansion using the approximation of \cite{lam} (and discussed in Appendix~\ref{reproducingST})
gives
\begin{eqnarray}
&&\hspace*{-5mm}\frac{\pa^3}{\pa B^3(S_n)}
\int_{-\infty}^{B_n} d\d_n\,
\left(\Pi^{(0)}_{\eps=0} +\Pi^{(1,\rm{ST})}_{\eps=0} +\Pi^{(2,\rm{ST})}_{\eps=0} +\cdots\right)\nonumber\\
&&=-\sqrt{\frac{2}{\pi S_n^3}}\left(1-\frac{B^2(S_n)}{S_n}+\frac{B_n}{S_n}{\cal P}(S_n)-
2\frac{{\cal P}^2(S_n)}{S_n}\right)\nonumber\\
&&\hspace*{5mm}\times e^{-B^2(S_n)/(2S_n)}\, . 
\label{pa3xc2}
\end{eqnarray}
where 
\begin{equation}
{\cal P}(S)\equiv\sum_{p=1}^5\frac{(-S)^p}{p!} \frac{\partial^p B(S)}{\partial S^p}\, .
\end{equation}
Notice that the sum runs only up to $p=5$  to provide a good fit to the data,
as mentioned earlier in Sect.~\ref{sect:STbarr}.
If we now  normalize the bispectrum as
\be\label{defbis}
{\cal S}_3(S)\equiv \frac{1}{S^2}\langle\d^3(S)\rangle\, ,
\ee
we finally obtain the leading NG contribution to the first-crossing rate
with a generic moving barrier.  Using (\ref{pa3xc}) we obtain 
\begin{eqnarray}
&&{\cal F}_{\rm NG}(S)=\mathcal{F}^{(0)}+\mathcal{F}^{(a)}+\mathcal{F}^{(b)}+\mathcal{F}^{(c)}
\nonumber\\
&&\hspace*{-7mm}
+\frac{\mathcal{S}_3}{12 \sqrt{2\pi } S^{5/2}} \left[
-2 \left(S^2+2 S B^2-B^4+S B B'\left(-7 S+B^2\right) \right.\right.\nn\\
&&\hspace*{-7mm}\left.\left.
-8 S^3 B'^2+4 S^3 B B'^3\right)+ S^3 B''\left(B+22 S B' \right)\right] 
e^{-B^2/(2S)}\nn\\
&&\hspace*{-7mm}
+\frac{S^2\mathcal{S}_3'}{3 \sqrt{2\pi } S^{5/2}} \left[
B^2+S B B'+S \left(-1+2 S B'^2\right)\right]e^{-B^2/(2S)}\nn\\
&&\hspace*{-7mm}
-{S\over 4}  
 \left(\left(2 \mathcal{S}_3+S \mathcal{S}_3'\right) B''+S \mathcal{S}_3 B'''\right)
 \textrm{Erfc}\left[\frac{B}{\sqrt{2S} }\right]\,, 
 \label{totalNG}
\end{eqnarray}
while  using (\ref{pa3xc2}) we obtain
\begin{eqnarray}
&&{\cal F}_{\rm NG}(S)={B+\mathcal{P}\over \sqrt{2\pi} S^{3/2}}e^{-B^2/(2S)}
\nn\\
&&\hspace*{-5mm}+\frac{{\cal S}_3}{6\sqrt{2\pi }S^{5/2}}\left[B^4-B^3({\cal P}+2S B')
+2B^2(-S+{\cal P}^2\right.\nonumber\\
&&\hspace*{-5mm}+S{\cal P}B')+ S B({\cal P}+6S B'-4{\cal P}^2B'
-2S{\cal P}')\nn\\
&&\hspace*{-5mm}\left.
-S(S+2{\cal P}({\cal P}+S B'-4S{\cal P}'))\right]e^{-B^2/(2S)}
\nonumber\\
&&\hspace*{-5mm}+\frac{S^2{\cal S}'_3}{3\sqrt{2\pi }S^{5/2}}\left[ B^2-B{\cal P}-S
+2{\cal P}^2\right]
e^{-B^2/(2S)}\, ,
\label{totalNGapprox}
\end{eqnarray}
where   the prime denotes differentation with respect to $S$. 

Both formualae (\ref{totalNG})-(\ref{totalNGapprox}) can be further improved using a saddle-point
technique in order to resum the largest contributions from NG, as in
\cite{D'Amico:2010ta}. Limiting this procedure to the 
leading terms of (\ref{totalNGapprox}) and treating $\mathcal{P}(S)$ and the derivatives of $B(S)$ as small parameters,
we find for instance
\begin{eqnarray}
&&{\cal F}_{\rm NG}(S)=
\frac{B e^{-{B^2\over 2S}}}{\sqrt{2\pi}S^{3/2}}
e^{{1\over 6}\mathcal{S}_3 {B^3\over S}}\left(1-{1\over 3}\mathcal{S}_3 B-
{1\over 6}{S\mathcal{S}_3\over B}\right)\nn\\
&&\hspace*{-5mm}+{\mathcal{P}\over \sqrt{2\pi} S^{3/2}}e^{-B^2/(2S)}
\nn\\
&&\hspace*{-5mm}+\frac{{\cal S}_3}{6\sqrt{2\pi }S^{5/2}}\left[-B^3({\cal P}+2S B')
+2B^2({\cal P}^2\right.\nonumber\\
&&\hspace*{-5mm}+S{\cal P}B')+ S B({\cal P}+6S B'-4{\cal P}^2B'
-2S{\cal P}')\nn\\
&&\hspace*{-5mm}\left.
-2S{\cal P}({\cal P}+S B'-4S{\cal P}')\right]e^{-B^2/(2S)}
\nonumber\\
&&\hspace*{-5mm}+\frac{S^2{\cal S}'_3}{3\sqrt{2\pi }S^{5/2}}\left[ B^2-B{\cal P}-S
+2{\cal P}^2\right]
e^{-B^2/(2S)}\,.
\label{totalNGexp}
\end{eqnarray}
Notice that,  in the limit of constant barrier, our formulae are slightly different from those of  \cite{D'Amico:2010ta}; we believe that
the origin of this difference is due to the fact that they assumed a very specific form for  the cumulants 
$\langle\delta_i\delta_j\delta_k\rangle\propto (S_i S_j S_k)^{1/2}$. With this assumption,  one can find relations between the various derivatives of the cumulants, which otherwise    are independent.

\begin{figure}
\centering
\includegraphics[width=0.45\textwidth]{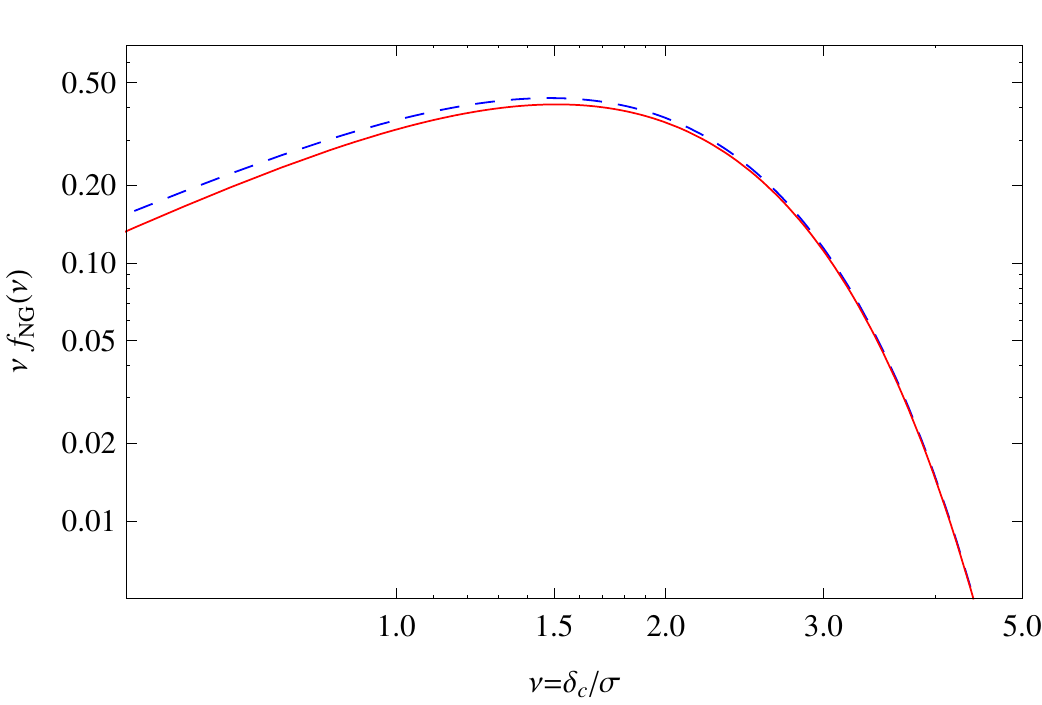}
\caption{
\label{figure:compareFNG}
 The first-crossing rate deduced from   Eqs.~(\ref{totalNG}) (dashed blue line) and (\ref{totalNGapprox}) (solid red line), for the case of ellipsoidal barrier (\ref{B(S)2}). 
We used  $\mathcal{S}_3$  given by Eq.~(\ref{S3fit}) with local $f_{\rm NL}= 100$.
}
\end{figure}

\begin{figure}
\centering
\includegraphics[width=0.45\textwidth]{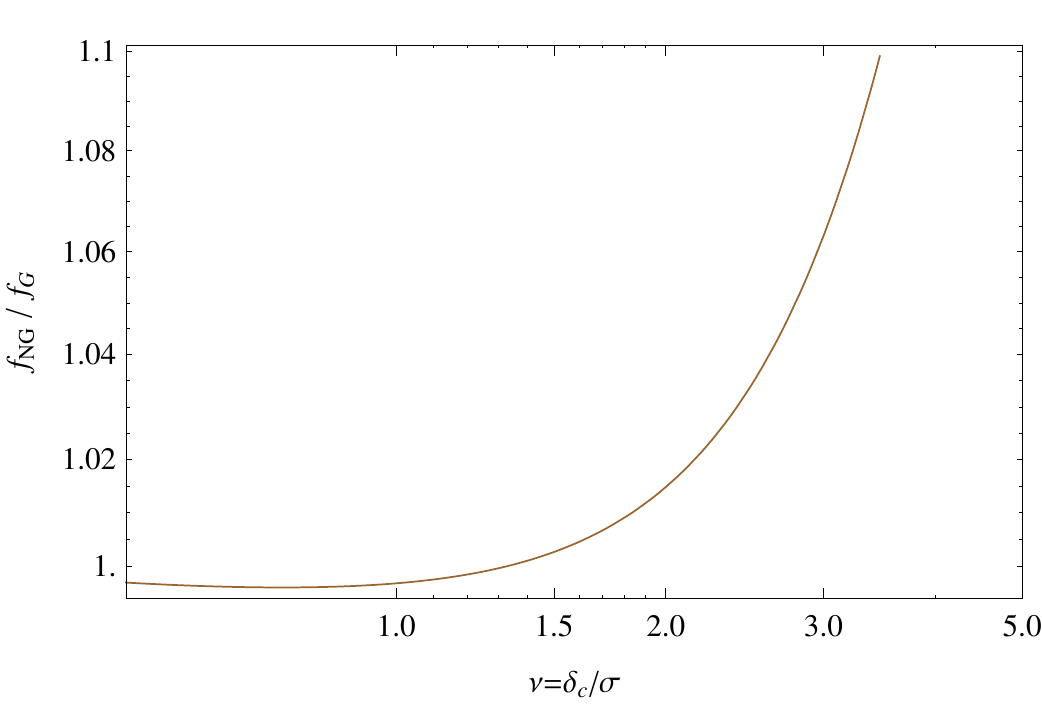}
\caption{
\label{figure:ratio}
 The ratio between the non-Gaussian first-crossing rate $f_{\rm NG}$ deduced from   Eqs.~(\ref{totalNGapprox}) and
 the Gaussian one $f_{\rm G}$.
We used  $\mathcal{S}_3$  given by Eq.~(\ref{S3fit}) with local $f_{\rm NL}= 100$.
}
\end{figure}

\begin{figure}
\centering
\includegraphics[width=0.45\textwidth]{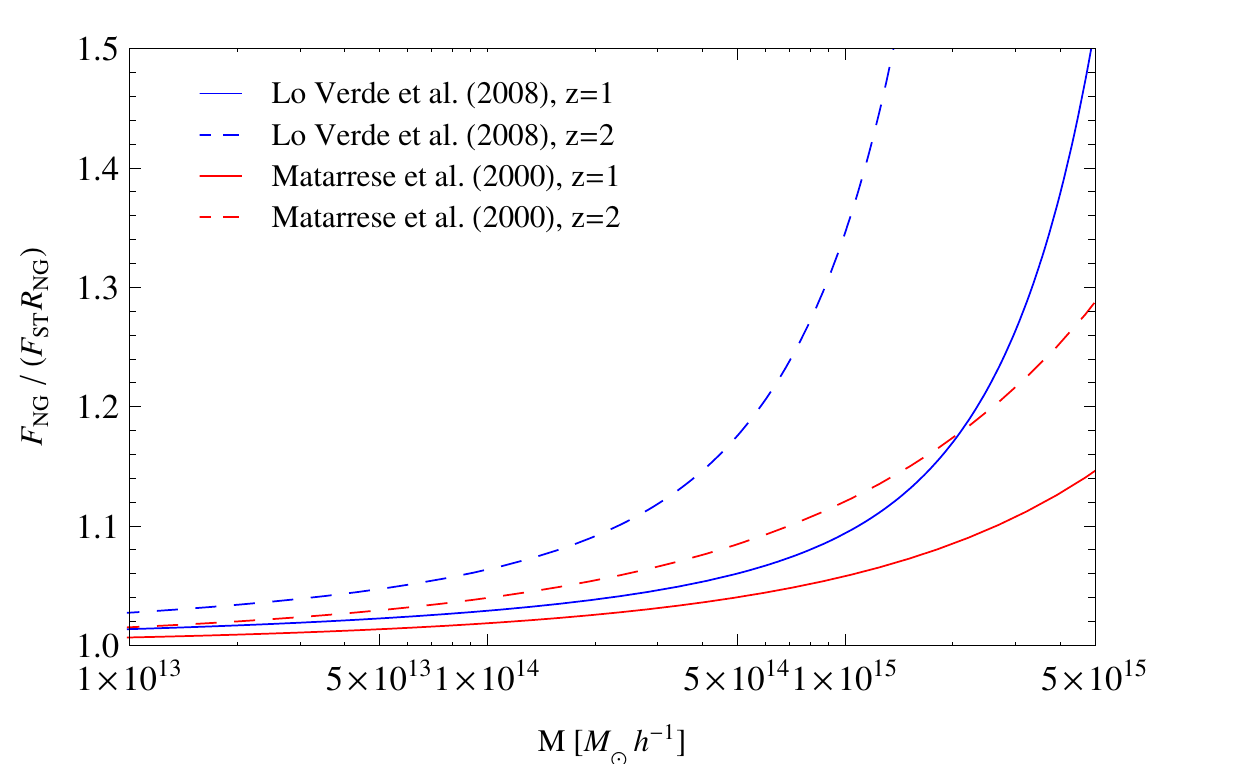}
\caption{\label{figure:ratioNG}
Ratio of the ${\cal F}_{\rm NG}(S)$ in (\ref{totalNGexp}) to the first-crossing rate given by the ${\cal F}_{\rm ST}(S)$ in (\ref{FST}) times a form factor 
$\mathcal{R}_{\rm NG}$, as a function of the halo mass $M$  for $f_{\rm NL}=100$.
 The form factors are those in Eq.~(\ref{RMat}) (red lines) and Eq.~(\ref{RLV}) (blue lines).
We considered redshifts $z=1$ (solid lines) and $z=2$ (dashed lines). }
\end{figure}

In the limit of constant barrier $B(S)=\sqrt{a}\delta_c$ one recovers the spherical collapse result of MR3
(neglecting the terms proportional to ${\cal S}_3'$)
\begin{eqnarray}
{\cal F}^{\rm sph}_{\rm NG}(S)&=&\frac{\sqrt{a}\delta_c}{\sqrt{2\pi}S^{3/2}}e^{-a\delta_c^2/(2S)}
\bigg[1+\nn\\
&&\left.\frac{S\,{\cal S}_3}{6\sqrt{a}\delta_c}\left(\frac{(\sqrt{a}\delta_c)^4}{S^{2}}-2\frac{(\sqrt{a}\delta_c)^2}{S}-1\right)\right]\, .
\label{FNGsph}
\end{eqnarray}
In Figure \ref{figure:compareFNG} we show the first-crossing rates (\ref{totalNG}) and (\ref{totalNGapprox}),
applied to the case of the ellipsoidal barrier (\ref{B(S)2}). The two curves differ by ${\cal O}(10)$\% at most in the small halo mass regime. In Figure \ref{figure:ratio} we plot the 
 ratio between the non-Gaussian first-crossing rate deduced from   Eqs.~(\ref{totalNG}) and
 the Gaussian one.
In Figure \ref{figure:ratioNG}  we show the ratios between the first-crossing rate given in (\ref{totalNGexp})
and the first-crossing rates (\ref{FSTNG}) built up from two different commonly used
 form factors $\mathcal{R}_{\rm NG}$, the one of \cite{MVJ}:
\begin{eqnarray}
\mathcal{R}_{\rm NG}&=&\exp\left[ {\mathcal{S}_3(\sqrt{a}\delta_c)^3\over  6S}\right]
\left[\sqrt{1-{1\over 3}(\sqrt{a}\delta_c)\mathcal{S}_3}\right.\nn\\
&&\left. +{1\over 6}{(\sqrt{a}\delta_c)^2\over
\sqrt{1-{1\over 3}(\sqrt{a}\delta_c)\mathcal{S}_3}}{d\mathcal{S}_3\over d\ln \sqrt{S}}\right]\,,
\nn\\
\label{RMat}
\end{eqnarray}
and the one of  \cite{LV}:
\begin{eqnarray}
\mathcal{R}_{\rm NG}&=&1+{1\over 6}{S\over \sqrt{a}\d_c}
\left[\mathcal{S}_3\left({(\sqrt{a}\delta_c)^4\over S^2}-2{(\sqrt{a}\delta_c)^2\over S}-1\right)\right.\nn\\
&&\left.+{d\mathcal{S}_3\over d\ln \sqrt{S}}\left({(\sqrt{a}\delta_c)^2\over S}-1\right)\right]\,.
\label{RLV}
\end{eqnarray}
In the plots we used  the conversion from the variable $S$ to the variable $M$ given in 
eq.~(A2) of \cite{Neistein}, while for the  scale-dependence of $\mathcal{S}_3$ we used the following simple fitting formula
\be
\mathcal{S}_3(S)={2.4 \times 10^{-4}\over S^{0.45}} f_{\rm NL}\,,
\label{S3fit}
\ee
which agrees well with \cite{LV}.

As we can see, the  first-crossing rate in the case of an ellipsoidal collapse
and when NG is present is not generically given by the Gaussian first-crossing rate for the ellipsoidal model multiplied
by the form factor obtained from the PS approach and can differ significantly from it  by ${\cal O}(10-50)$\% or more at high redshift and large halo masses. 

\section{Conclusions}
\vspace{5mm}
Excursion set theory provides an elegant analytical 
technique to describe the distribution of dark matter in our universe.
When supplemented with various improvement concerning the physical modelisation 
of halo formation (such as the ellipsoidal barrier of \cite{ST} to take into account the triaxiality of halo collapse and the diffusing barrier of MR2 to take into account the stochasticity inherent to the process), as well as with improvements on some technical aspects (such as the inclusion of the 
non-Markovian dynamics introduced by the filter function), it
provides a quantitative agreement with N-body simulations at the level of about 10\% in most of the interesting mass range. While even more accurate results might be needed for precision cosmology,
it is still remarkable that such a relatively simple theory catches quantitatively a significant part of the physics of such a complicated dynamical process as the formation of dark matter halos.
The same is true if the excursion set method is applied to describe the abundances of cosmic sheets and filaments. In this paper we have
extended  the  path integral approach proposed in MR1 for the spherical collapse case
to the case of generic moving barriers using a top hat window function in wavenumber space . We have shown that, using a well controlled and systematic expansion, we can reproduce the ST halo mass function very well, therefore putting it
on firmer grounds. We have also performed the computation of the first-crossing rate
 for the ellipsoidal barrier in the presence of non-Gaussian initial conditions. Our result
is given in  eq.~(\ref{totalNGexp}): it  is fully consistent in the sense that it does not require the introduction of any form factor artificially obtained from the PS formalism based on the spherical collapse
and in fact it provides a halo mass function which quantitatively differs from the one obtained from 
the form factor procedure.

\vspace{5mm}
\noindent
\paragraph*{Acknowledgements.}
We thank Ravi Sheth and Ruth Durrer for useful discussions.
The work of ADS is supported  in part 
by the U.S.  Department of Energy  under contract No. 
DE-FG02-05ER41360.  
The work of MM is supported by the Fond National Suisse. 
The work of AR is supported by 
the European Community's Research Training Networks 
under contract MRTN-CT-2006-035505.
\vspace{5mm}

\appendix

\section{Reproducing the first-crossing rate of Sheth \& Tormen}
\label{reproducingST}

We first compute $\Pi^{(1)}$. Using eqs. (\ref{byparts}), 
(\ref{Pigammafinal}) and (\ref{Pigammafinalbis}), 
the expression of  $\Pi^{(1)}_{\eps} (\d_n;S_n)$ in eq.~(\ref{Pi1}) can be rewritten as
\begin{eqnarray}
\Pi^{(1)}_{\eps=0} (\d_n;S_n)&=&\frac{B_n(B_n-\delta_n)}{\pi}\sum_{p=1}^\infty\frac{(-1)^p}{p!}\Bnp\label{ainApp}
\\
&&\times\int_0^{S_n}d S_i\frac{\left(S_n-S_i\right)^{p-(3/2)}
}{S_i^{3/2}}\nn\\
&&\times e^{-B_n^2/(2S_i)}e^{-(B_n-\delta_n)^2/[2 (S_n-S_i)]}\, .\nn
\end{eqnarray}
Instead of computing directly this integral, we now recall that to compute the first-crossing rate
(\ref{firstcrossmoving}) we need to compute the first derivative of $\Pi_{\eps} (\d_n;S_n)$ evaluated
at $\delta_n=B(S_n)$.  Since the  integral in eq. (\ref{ainApp}) is finite in the limit $\delta_n\rightarrow B(S_n)$,  taking  the approximation
$(S_n-S_i)^{p-1}\simeq (S_n)^{p-1}$ does not alter the convergence properties of the integral, but simplifies significantly its computation. This is equivalent to the approximation made by
\cite{lam}, see in particular the discussion below their eq.~(20).
Exploiting the fact that
\begin{eqnarray}
&&\int_0^{S_n}d S_i\frac{1}{S_i^{3/2}(S_n-S_i)^{1/2}}\nn\\
&&\times e^{-B^2(S_n)/(2S_i)} e^{-(B(S_n)-\delta_n)^2/(2 (S_n-S_i))}
\nonumber\\
&=&\frac{\sqrt{2\pi}}{B(S_n)}\frac{1}{S_n^{1/2}}{\rm exp}\left\{-\frac{(2B(S_n)-\delta_n)^2}{2 S_n}\right\}\, ,
\label{eq108}
\end{eqnarray}
we find  that  $\Pi^{(1, \rm{ST})}_{\eps=0} (\d_n;S_n)$ (where the superscript reminds us that
we have approximated the integral) is given by
\begin{eqnarray}
\label{zz}
\Pi^{(1, \rm{ST})}_{\eps=0} (\d_n;S_n)&=&
{2(B(S_n)-\delta_n)\over \sqrt{2\pi} S_n^{3/2}}
 e^{-(2B(S_n)-\delta_n)^2/(2 S_n)}\nn\\
&&\times \sum_{p=1}^\infty\frac{(-S_n)^p}{p!}B_n^{(p)}\, .
\end{eqnarray}
Next, we compute $\Pi^{(2)}_{\eps} (\d_n;S_n)$.
The  sum  over $i,j$  in eq.~(\ref{s2})  can be split into a sum over $i=j$ and a sum over $i<j$. The former does not contain
a finite part in the continuum limit and its divergence cancels against the divergent part of the 
latter sum (see appendix~B of MR1). Thus, we are reduced to compute the finite part of the sum over $i<j$. Proceeding as before
for the calculation of $\Pi^{(1)}_{\eps} (\d_n;S_n)$, and taking again $(S_n -S_i)^{p-1}\simeq S_n^{p-1}$
we obtain
\begin{eqnarray}
\label{s2intermediate}
&&\Pi^{(2, \rm{ST})}_{\eps=0} (\d_n;S_n)={B(S_n)(B(S_n)-\delta_n)\over \pi\sqrt{2\pi}}
\nn\\
&&
\phantom{=}\times  \sum_{p,q=1}^\infty\frac{B_n^{(p)}}{p!}
\frac{B_n^{(q)}}{q!}\left(-S_n\right)^{p-1}
\left(-S_n\right)^{q-1}\nn\\
&&
\phantom{=}\times\int_0^{S_n}d S_i {(S_n-S_i) e^{-{B^2(S_n)\over 2S_i}}\over S_i^{3/2}}  \nn\\
&&\phantom{=} \times \int_{S_i}^{S_n}d S_j { e^{-(B(S_n)-\delta_n)^2/(2(S_n-S_j))}\over (S_j-S_i)^{3/2}
(S_n-S_j)^{1/2}}  \, .
\end{eqnarray}
Let us indicate the integral by ${\cal A}(\d_n, S_n)$. It is convenient to perform the inner
 integral by deriving with respect to $\delta_n$
\begin{eqnarray}
&&\partial_n {\cal A}(\delta_n, S_n)=
\int_0^{S_n}d S_i {(S_n-S_i) e^{-{B^2(S_n)\over 2S_i}}\over S_i^{3/2}}\nn\\
&&\phantom{=}\times \int_{S_i}^{S_n}d S_j{(B(S_n)-\d_n) e^{-{(B(S_n)-\delta_n)^2\over 2(S_n-S_j)}}\over (S_j-S_i)^{3/2}
(S_n-S_j)^{3/2}}  \nn\\
&&=
\sqrt{2\pi}\int_0^{S_n}d S_i {e^{-{B^2(S_n)\over 2S_i}-{(B(S_n)-\delta_n)^2\over 2(S_n-S_i)}}
\over S_i^{3/2}(S_n-S_i)^{1/2}}\nn\\
&&\phantom{=}\times\left[1-{(B(S_n)-\delta_n)^2\over S_n-S_i}\right]
\nn\\
&=&{2\pi\over B(S_n)}{1\over S_n^{1/2}} e^{-(2B(S_n)-\delta_n)^2 /(2S_n) }\nn\\
&&\times\left[1-{(2B(S_n)-\d_n)(B(S_n)-\d_n)\over S_n}
\right]\,,
\end{eqnarray}
 where we  used eq.~(B.26)  of MR1 in the second line and eqs.~(A.5) of MR1 and (\ref{eq108})  in the third line. Integrating over $\delta_n$ we find
\be
{\cal A}(\delta_n, S_n)=-{2\pi\over S_n^{1/2}} {(B(S_n)-\d_n)\over B(S_n)} e^{-(2B(S_n)-\delta_n)^2/(2 S_n)}\,,
\ee 
 which can then be inserted into eq.~(\ref{s2intermediate}) to give
\begin{eqnarray}
&&\Pi^{(2, \rm{ST})}_{\eps=0} (\d_n;S_n)=
-{2(B(S_n)-\delta_n)^2\over \sqrt{2\pi} S_n^{5/2}}  \nn\\
&&\times e^{-(2B(S_n)-\delta_n)^2/(2 S_n)}\left[\sum_{p=1}^\infty\frac{(-S_n)^p}{p!}B_n^{(p)}
\right]^2 \,.
\end{eqnarray}
A similar procedure can be used to show that higher order contributions $\Pi^{(n, \rm{ST})}_{\eps=0}$ $(n>2$)
vanish as $(B(S_n)-\delta_n)^n$ when $\delta_n$ approches the barrier value $B(S_n)$.

The calculation of the first-crossing rate is then straightforward,  through  eq. (\ref{firstcrossmoving}). The zero-th order contribution from $\Pi^{(0)}_{\eps=0}$ is given by eq.~(\ref{F0}),
while the first-order contribution from $\Pi^{(1, \rm{ST})}_{\eps=0}$ reads
\begin{eqnarray}
{\cal F}^{(1, \rm{ST})}(S)=
\frac{1}{\sqrt{2\pi}S^{3/2}}e^{-B^2(S)/(2S)}
\sum_{p=1}^\infty\frac{(-S)^p}{p!}  \frac{\partial^p B(S)}{\partial S^p}\,
\end{eqnarray}
Higher-order contributions to the first-crossing rate vanish. This is already clear from the  contribution arising from the second-order
$\Pi^{(2, \rm{ST})}_{\eps=0}$
\begin{eqnarray}
{\cal F}^{(2, \rm{ST})}(S_n)=-\left[\sum_{p=1}^\infty\frac{(-S_n)^p}{p!}\frac{\partial^p B_n}{\partial S_n^p}
\right]^2 {e^{-(2B_n-\delta_n)^2/(2 S_n)}\over \sqrt{2\pi} S_n^{7/2}}  \nn\\
\times (B_n-\d_n)(3B_n\delta_n+2S_n-2B^2_n-\delta_n^2)
\,,
\end{eqnarray}
which vanishes for $\delta_n= B_n$.
The total first-crossing rate for a moving barrier, in the approximation discussed above, 
is therefore given by 
\begin{eqnarray}
{\cal F}_{\rm{ST}}(S)&=&\frac{e^{-B^2(S)/(2S)}}{\sqrt{2\pi}S^{3/2}}
\sum_{p=0}^\infty\frac{(-S)^p}{p!} \frac{\partial^p B(S)}{\partial S^p}\, .
\end{eqnarray}

\section{Computation of $\Pi_{\eps=0}^{(a)}$, $\Pi_{\eps=0}^{(b)}$, $\Pi_{\eps=0}^{(c)}$}
\label{App:A}

In this appendix we compute the contribution to $\Pi_{\eps=0}$ in the derivative expansion discussed in Section~\ref{sect:der}. 
The first, using the techniques discussed in MR1, is simply computed,
\bees
&&\hspace*{-5mm}\Pi^{(a)}_{\eps=0}(\d_n;S_n)= -\frac{1}{\pi}\frac{dB_n}{dS_n} B_n (B_n-\d_n)\nn\\
&&\hspace*{-5mm}\times\int_0^{S_n} dS_i\, \frac{1}{S_i^{3/2}(S_n-S_i)^{1/2}}
\exp\left\{ -\frac{B_n^2}{2 S_i}-\frac{(B_n-\d_n)^2}{2(S_n-S_i)}\right\}\nn\\
&&\hspace*{-5mm}= -\(\frac{2}{\pi}\)^{1/2} \frac{dB_n}{dS_n}\, \frac{(B_n-\d_n)}{S_n^{1/2}}
\exp\left\{ -\frac{(2B_n-\d_n)^2}{2S_n}\right\}\, .
\ees
The second term is 
\bees
&&\hspace*{-5mm}
\Pi^{(b)}_{\eps=0}(\d_n;S_n)= \frac{1}{2\pi}\frac{d^2B_n}{dS^2_n} B_n (B_n-\d_n)\nn\\
&&\hspace*{-5mm}
\times\int_0^{S_n} dS_i\, \frac{(S_n-S_i)^{1/2}}{S_i^{3/2}}
\exp\left\{ -\frac{B_n^2}{2 S_i}-\frac{(B_n-\d_n)^2}{2(S_n-S_i)}\right\}\nn\\
&&\hspace*{-5mm}
= \frac{1}{2\pi} \frac{d^2B_n}{dS^2_n} (B_n-\d_n)\, \\
&&\hspace*{-5mm}
\times\[ \sqrt{2\pi} S_n^{1/2}e^{-(2B_n-\d_n)^2/(2S_n)}
-\pi B_n {\rm Erfc}\(\frac{2B_n-\d_n}{\sqrt{2S_n}}\)\]\, ,\nn
\ees
where the integral has been computed using eq.~(109) of MR1. The last term is the most complicated. Using the 
$\a$-regularization and the finite part prescription developed in  Appenix.~B of MR1, we find as usual that the terms in the sum with $i=j$ have a vanishing finite part, while the contribution from the terms with $i<j$ (plus an equal contribution from $i>j$) can be written as
\bees
&&\hspace*{-5mm}\Pi^{(c)}_{\eps=0}(\d_n;S_n)=
 \frac{B_n(B_n-\d_n)}{\pi\sqrt{2\pi}} \(\frac{dB_n}{dS_n}\)^2\nn\\
&&\hspace*{-5mm}\times{\cal FP}
\int_0^{S_n}dS_i\int_{S_i}^{S_n} dS_j\, 
\frac{(S_n-S_i)}{S_i^{3/2} (S_j-S_i)^{3/2} (S_n-S_j)^{1/2}}\nn\\
&&\hspace*{-5mm}\times\exp\left\{ -\frac{B_n^2}{2 S_i}-\frac{\a\eps}{2(S_j-S_i)}
-\frac{(B_n-\d_n)^2}{2(S_n-S_j)}\right\}\nn\\
&&\hspace*{-5mm}=
 \frac{B_n(B_n-\d_n)}{\pi\sqrt{2\pi}} \(\frac{dB_n}{dS_n}\)^2
\int_0^{S_n}dS_i\, \frac{(S_n-S_i)}{S_i^{3/2} }\\
&&\hspace*{-5mm}\times \exp\left\{ -\frac{B_n^2}{2 S_i}\right\}  {\cal FP}\int_{S_i}^{S_n} dS_j\, 
\frac{1}{(S_j-S_i)^{3/2} (S_n-S_j)^{1/2}}\,\nn\\
&&\hspace*{-5mm}\times\exp\left\{ -\frac{\a\eps}{2(S_j-S_i)}
-\frac{(B_n-\d_n)^2}{2(S_n-S_j)}\right\}\, ,
\label{(c)}
\ees
where ${\cal FP}$ denotes the finite-part prescription developed in App.~B of MR1.
The integral over $dS_j$ is performed using MR1, eq.~(108), and is equal to
\be
\frac{\sqrt{2\pi}}{\sqrt{\a\eps}}\, \frac{1}{(S_n-S_i)^{1/2}}
\exp\left\{ -\frac{(B_n-\d_n+\sqrt{\a\eps})^2}{2(S_n-S_i)}\right\}\, .
\ee
Expanding the exponential we therefore get a singularity $1/\sqrt{\eps}$ (which is canceled by a similar singularity in the term of the sum with $i=j$, see MR1), and a finite part, given by
\be
-\sqrt{2\pi} \frac{(B_n-\d_n)}{(S_n-S_i)^{3/2}}\, e^{-(B_n-\d_n)^2/[2(S_n-S_i)]}\, .
\ee
The remaining integral over $dS_i$ is performed again using MR1, eq.~(108), so finally
\bees
\Pi^{(c)}_{\eps=0}(\d_n;S_n)&=& 
-\(\frac{2}{\pi}\)^{1/2}
(B_n-\d_n)^2\(\frac{dB_n}{dS_n}\)^2\,\frac{1}{S_n^{1/2}}\nn\\
&&\times\exp\left\{ -\frac{(2B_n-\d_n)^2}{2S_n}\right\}\, .
\ees

\section{Computation of $\Pi_{\eps=0}^{(1)}$}\label{App:B}

In this appendix we fill the missing step in the computation of
$\Pi_{\eps=0}^{(1)}$. The issue is the computation of
the integral
\bees
{\cal I}_p(a,b, S_n)&\equiv&\int_0^{S_n} dS_i\,  S_i^{-3/2} (S_n-S_i)^{p-\frac{3}{2}}\nn\\
&&\times \exp\left\{ -\frac{a^2}{2S_i}-\frac{b^2}{2(S_n-S_i)}\right\}\, ,
\ees
where $a\equiv B_n>0$ and 
$b\equiv (B_n-\d_n)>0$. Changing the integration 
variable to $z=(S_n/S_i)-1$ we get
\bees
{\cal I}_p (a,b, S_n)&=&S_n^{p-2}\exp\left\{ -\frac{a^2+b^2}{2S_n}\right\}
\int_0^{\infty}dz\nn\\ 
&&\times\( \frac{1}{z^{3/2}}+ \frac{1}{z^{1/2}}\)\(\frac{z}{1+z}\)^p\nn\\
&&\times 
\exp\left\{ -\(\frac{a^2}{2S_n}\)z-\(\frac{b^2}{2S_n}\)\, \frac{1}{z}
\right\}\,.
\ees
For $p=0,1$ the integral can be performed exactly (see eq.~9.471.12 of \cite{Grad})
and we get\footnote{These integrals were already computed exactly in a different way in MR1. We thank Ruth Durrer for suggesting this more direct derivation.}
\bees
{\cal I}_0 (a,b, S_n)&=&\frac{(2\pi)^{1/2}}{S_n^{3/2}}\, \frac{a+b}{ab}\, e^{-(a+b)^2/(2S_n)}
\, ,\label{I0ab}\\
{\cal I}_1 (a,b, S_n)&=&\frac{(2\pi)^{1/2}}{S_n^{1/2}}\, \frac{1}{a}\, e^{-(a+b)^2/(2S_n)}\, .
\ees
For $p\geq 2$  we have not been able to compute the integral exactly. 
However, as discussed in the text, for computing the first-crossing rate it is sufficient to evaluate it at $b=0$. The resulting integral can be computed (e.g. using Mathematica) in terms of the
confluent hypergeometric function $U(a,b,z)$,
\bees
{\cal I}_p (a,0, S_n)&=&S_n^{p-2}\, \frac{\sqrt{2S_n}}{a}\,  e^{-a^2/(2S_n)}\nn\\
&&\times \Gamma\(p-\frac{1}{2}\) U\(p-1,\frac{1}{2},\frac{a^2}{2S_n}\)\, .
\label{Ipa0}
\ees
Observe that $U(0,b,z)=1$ and $\Gamma(1/2)=\sqrt{\pi}$, so \eq{Ipa0} also reproduces
correctly ${\cal I}_p (a,0, S_n)$ when $p=1$.
It is also useful the limit
\begin{eqnarray}
{\cal I}_p (0,0, S_n)&\equiv& \mathcal{FP}\lim_{a\to 0}{\cal I}_p (a,0, S_n)
=-\pi c_p S_n^{p-2}\,,
\end{eqnarray}
where the coefficients $c_p$ are given by
\be
c_p={2\over \sqrt{\pi}}{\Gamma\left(p-{1\over 2}\right)\over \Gamma\left(p-1\right)}\,.
\label{cp}
\ee

\section{Computation of the general term $\Pi_{\eps=0}^{(m)}$ in the limit $(B_n-\d_n)\ra 0$}\label{App:C}

The general term $\Pi_{\epsilon=0}^{(m)}$ is given by
\begin{eqnarray}
\Pi_{\epsilon=0}^{(m)}&=&{1\over m!}\sum_{p_1,\ldots,p_m=1}^\infty {B_n^{(p_1)}\cdots B_n^{(p_m)}
\over p_1!\cdots p_m!}\nn\\
&&\times\sum_{i_1,\ldots, i_m=1}^{n-1}
(S_{i_1}-S_n)^{p_1}\cdots (S_{i_m}-S_n)^{p_m}\nn\\
&&\times
\int_{-\infty}^{B_n}d\delta_1\cdots d\delta_{n-1}\partial_{i_1}\cdots \partial_{i_m} W^{\rm gm}\,.
\label{Pimdef}
\end{eqnarray}
The last integral is equal to
\begin{eqnarray}
&\int_{-\infty}^{B_n}d\delta_1\cdots d\delta_{n-1}\partial_{i_1}\cdots \partial_{i_m}  W^{\rm gm}=&\nn\\
&\Pi^{\rm gm}(\delta_0, B_n, S_{i_1})
 \Pi^{\rm gm}(B_n, B_n, S_{i_2}-S_{i_1})\cdots&\nn\\
&\cdots
 \Pi^{\rm gm}(B_n, B_n, S_{i_m}-S_{i_{m-1}})\Pi^{\rm gm}(B_n, \delta_n, S_n-S_{i_m}).&
\nn
\end{eqnarray}
Using eqs.~(\ref{Pigammafinal})-(\ref{Pigammafinalter}) for $\Pi^{\rm gm}$, eq.~(\ref{Pimdef}) becomes
\begin{eqnarray}
\Pi_{\epsilon=0}^{(m)}&=&{1\over m!}{B_n(B_n-\delta_n)\over 2^{m-1\over 2}\pi^{m+1\over 2}}\sum_{p_1,\ldots,p_m=1}^\infty (-1)^{p_1+\ldots+ p_m}\nn\\
&& \times {B_n^{(p_1)}\cdots B_n^{(p_m)}
\over p_1!\cdots p_m!}
\mathcal{J}^{(m)}_{p_1,\ldots,p_m}(B_n, S_n)\nn\\
&&+\mathcal{O}(B_n-\delta_n)^2\,,
\label{Pim2}
\end{eqnarray}
where
\begin{eqnarray}
&&\mathcal{J}_{p_1,\ldots,p_m}^{(m)}(B_n, S_n)\equiv\mathcal{FP}\int_0^{S_n}dS_{i_1} {(S_n-S_{i_1})^{p_1}\over S_{i_1}^{3/2}} e^{-{B_n^2\over 2S_{i_1}}}\nn\\
&&\hspace*{15mm} \times\int_{S_{i_1}}^{S_n}dS_{i_2} {(S_n-S_{i_2})^{p_2}
\over (S_{i_2}-S_{i_1})^{3/2}}\times ( \ldots)\nn\\
&&\hspace*{15mm}\times
\int_{S_{i_{m-1}}}^{S_n}dS_{i_m} {(S_n-S_{i_m})^{p_m-3/2}\over (S_{i_m}-S_{i_{m-1}})^{3/2}} \,.
\end{eqnarray}

{We have only considered the finite parts from the
sum with $i_1<i_2<\cdots <i_m$, because the divergent parts all cancel. A priori,
we cannot exclude that there may be other finite contributions to the sum coming from
terms with $i_1<\cdots <i_k=i_{k+1}<\cdots i_m$. However, we expect the contribution
we compute here as representative of the correct result. 
}

The integral $\mathcal{J}_{p_1,\ldots,p_m}^{(m)}(B_n, S_n)$ satisfies the recursion
relation
\begin{eqnarray}
\mathcal{J}_{p_1,\ldots,p_{m}}^{(m)}(B_n, S_n)&=&
\int_{0}^{S_n}d S_{i} {e^{-B_n^2/(2S_i)}(S_n-S_i)^{p_1}\over S_i^{3/2}}\nn\\
&&\times \mathcal{J}_{p_2,\ldots,p_{m}}^{(m-1)}(0, S_n-S_{i})\,.
\label{recursion}
\end{eqnarray}
Let us set
\be
\mathcal{J}_{p_1,\ldots, p_{m}}^{(m)}(0, y)=(-\pi)^m c_{p_1,\ldots, p_m} \,
y^{p_1+\ldots+p_m-{m+3\over 2}}\,,
\label{ansatz}
\ee
where the coefficients $c$ are now to be determined. 
We insert the ansatz above into the recursion relation  (\ref{recursion}) for
 $\mathcal{J}_{p_1,\ldots,p_{m}}^{(m+1)}(B_n, S_n)$ and obtain
\begin{eqnarray}
&\mathcal{J}_{p_1,\ldots,p_{m+1}}^{(m+1)}(B_n, S_n)=(-\pi)^m c_{p_2,\ldots,p_{m+1}}&\nn\\
&\times\int_{0}^{S_n}d S_{i} {(S_n-S_i)^{p_1+\ldots+p_{m+1}-{m+3\over 2}}}S_i^{-3/2}
e^{-B_n^2/(2S_i)}&
\label{Im1}
\,.
\end{eqnarray}
The previous integral is solved with the substitution $z=(S_n/S_i)-1$ and it evaluates to
\begin{eqnarray}
&S_n^{\sum_{k=1}^{m+1}p_k-{m\over 2}-2}\int_0^\infty dz {z^{\sum_{k=1}^{m+1}p_k-{m+3\over 2}}\over (1+z)^{\sum_{k=1}^{m+1}p_k-{m\over 2}-1}}e^{-{B_n^2\over 2S_n}(1+z)}&\nn\\
&=S_n^{\sum_{k=1}^{m+1}p_k-{m\over 2}-2}
e^{-{B_n^2\over 2S_n}}\sqrt{{2S_n\over B_n^2}}\,
\Gamma\left(\sum_{k=1}^{m+1}p_k-{m+1\over 2}\right)&
\nn\\
&\times
U\left(\sum_{k=1}^{m+1}p_k-{m+1\over 2},{1\over 2}, {B_n^2\over 2S_n}\right),&
\end{eqnarray}
therefore  eq.~(\ref{Im1}) becomes
\begin{eqnarray}
&\mathcal{J}_{p_1,\ldots,p_{m+1}}^{(m+1)}(B_n, S_n)=
(-\pi)^m c_{p_2,\ldots,p_{m+1}}S_n^{\sum_{k=1}^{m+1}p_k-{m\over 2}-2}&\nn\\
&\times 
\sqrt{{2S_n\over B_n^2}}e^{-{B_n^2\over  2S_n}}\Gamma\left(\sum_{k=1}^{m+1}p_k-{m+1\over 2}\right)
&\nn\\
&\times
U\left(\sum_{k=1}^{m+1}p_k-{m+1\over 2},{1\over 2}, {B_n^2\over 2S_n}\right)\,.&
\label{Im2}
\end{eqnarray}
We can evaulate eq.~(\ref{Im2}) in the limit $B_n^2/(2S_n)\to  0$, and retain 
the finite part only (as the divergent terms all cancel in the end):
\begin{eqnarray}
&\mathcal{J}_{p_1,\ldots,p_{m+1}}^{(m+1)}(0, y)=-2\sqrt{\pi}(-\pi)^m c_{p_2,\ldots,p_{m+1}}&
\nn\\
&\times{\Gamma\left(\sum_{k=1}^{m+1}p_k-{m\over 2}-{1\over 2}\right)\over 
\Gamma\left(\sum_{k=1}^{m+1}p_k-{m\over 2}-1\right)}
y^{\sum_{k=1}^{m+1}p_k-{m\over 2}-2}\,.&
\end{eqnarray}

On the other hand, the left-hand side of the previous relation can be expressed by
(\ref{ansatz}) and we then arrive at a recursion relation for the coefficients $c$ (after
relabelling $m\to m-1$ for convenience):
\be
c_{p_1,\ldots, p_m}={2\over \sqrt{\pi}}
{\Gamma\left(\sum_{k=1}^{m}p_k-{m\over 2}\right)\over 
\Gamma\left(\sum_{k=1}^{m}p_k-{m+1\over 2}\right)}
c_{p_2,\ldots, p_m}\,,
\label{cpm}
\ee
which is valid for $m\geq 2$,  while for $m=1$ we have already found in (\ref{cp})
\be
c_p={2\over \sqrt{\pi}}{\Gamma\left(p-{1\over 2}\right)\over \Gamma\left(p-1\right)}\,.
\label{cp2}
\ee
Equations (\ref{cpm})-(\ref{cp2}) define recursively  the coefficients $c$  and 
 it is possible to find them easily up to any desired order.
As the $c$ appear in the generic integral  (\ref{Im2}), which in turn appears in (\ref{Pim2}),
it is then possible to write down the result for the  generic term $\Pi^{(m)}$:
\begin{eqnarray}
\Pi_{\epsilon=0}^{(m)}&=&{(B_n-\delta_n)e^{-{B_n^2\over 2S_n}}\over m! \,2^{{m\over 2}-1}\pi^{3-m\over 2}}
\sum_{p_1,\ldots,p_m=1}^\infty (-1)^{\sum_{k=1}^m p_k+m+1}\nn\\
&\times&{B_n^{(p_1)}\cdots B_n^{(p_m)}
\over p_1!\cdots p_m!} c_{p_2,\ldots, p_m}S_n^{\sum_{k=1}^m p_k-{m\over 2}-1}\nn\\ 
&\times& \Gamma\left(\sum_{k=1}^m p_k-{m\over 2}\right)
U\left(\sum_{k=1}^m p_k-{m+1\over 2},{1\over 2}, {B_n^2\over 2S_n}\right)\nn\\
&+&\mathcal{O}(B_n-\delta_n)^2.
\end{eqnarray}

\bibliographystyle{mn2e}

\begin{thebibliography}{99}

\bibitem[Acquaviva {et al.}(2003)]{acquaviva}
Acquaviva, V.,  Bartolo, N., Matarrese, S.  and  Riotto, A. 2003,    
Nucl. Phys.  {B667}, 119. 

\bibitem[Afshordi \& Tolley(2008)]{tolley}
Afshordi, N. \& Tolley, A.   2008,  Phys. Rev. D78, 123507. 

\bibitem[Audit {et.~al.}(1997)]{Audit}
Audit, E., Teyssier, R. and Alimi, J.-M., 1997,
Astron. Astrophys. 325,439.

\bibitem[Bardeen {et~al.}(1986)]{BBKS}
 Bardeen, J.M., Bond,  J.R., Kaiser, N.   and  Szalay, A.S. 1986,
  ApJ {304}, 15.

\bibitem[Bartolo  {et~al.}(2004)]{bartoloreview}
Bartolo N., Komatsu E., Matarrese S. \& Riotto A.  2004,  Phys. Rept.  402, 103. 

\bibitem[Bartolo  {et~al.}(2005)]{bartolosig}
 Bartolo N.,  Matarrese S. \& Riotto A.  2005,  JCAP 0510,  010. 

\bibitem[Bond {et~al.}(1991)]{Bond}
  Bond, J.~R., Cole, S., Efstathiou, G. \& Kaiser, N. 1991,
  ApJ.  { 379},  440.

\bibitem[Bond \& Myers(1996)]{BondMyers} 
Bond, J.~R. and Myers, S. 1996, ApJS, 103, 1.

\bibitem[Carbone et al.(2008)]{CVM} Carbone C., Verde
L., Matarrese S., 2008, ApJ, 684, 1.

\bibitem[Dalal et al.(2008)]{Dalal} Dalal N., Dore' O.
Huterer D.,Shirokov A., 2008, Phys. Rev. D77, 123514.

\bibitem[D'Amico et al.(2010)]{D'Amico:2010ta}
  D'Amico, G., Musso, M., Norena, J. and Paranjape, A. 2010,
  arXiv:1005.1203 [astro-ph.CO].

\bibitem[Furlanetto et al. (2004)]{furl}
Furlanetto, S., Zaldarriaga, M. and Hernquist, L., 2004,  ApJ 613, 1.

\bibitem[Giannantonio \& Porciani(2010)]{porciani}
Giannantonio T. \&  Porciani C. 2010, Phys. Rev. D81, 063530.

\bibitem[Gradstein \& Ryzhik(1980)]{Grad}
Gradstein L. S.  and Ryzhik I. M. (1980), {\em Tables of Integrals, Series and Products}, Academic
Press, 1980.

\bibitem[Grinstein \& Wise(1986)]{GW} Grinstein B.
  \& Wise, M.~B.\ 1986, ApJ, 310, 19. 

\bibitem[Grossi et al.(2009)]{grossi2009} 
Grossi M.  et al., 2009, MNRAS { \bf 398}, 321.

\bibitem[Jenkins {et~al.}(2001)]{jenkins} Jenkins, A.  
{\em et al.} 2001, MNRAS { \bf 321}, 372.

\bibitem[Koyama et al.(1999)]{KOYAMA} Koyama, K., Soda,
J., \& Taruya, A.\ 1999, MNRAS, 310, 1111. 

\bibitem[Lam \& Sheth(2009)]{lam} Lam, T.Y.  \& Sheth, R. 2009, MNRAS, 398,214L.

\bibitem[Lee \& Shandarin(1998)]{LeeS} 
Lee, J. \& Shandarin,~S.~F. 1998, ApJ, 500, 14.

\bibitem[{LoVerde} {et~al.}(2008)]{LV}
LoVerde, M., Miller, A., Shandera,~S. \& Verde,~L. 2008,
JCAP 0804, 014.

\bibitem[Lucchin et al.(1988)]{LMV} Lucchin, F.,
Matarrese, 
S., \& Vittorio, N.\ 1988, ApJl, 330, L21. 

\bibitem[Maggiore \& Riotto(2010a)]{MR1} Maggiore, M.  \& Riotto, A. 2010a, ApJ, 711, 907.

\bibitem[Maggiore \& Riotto(2010b)]{MR2} Maggiore, M.  \& Riotto, A. 2010b, 
ApJ 717, 515.

\bibitem[Maggiore \& Riotto(2010c)]{MR3} Maggiore, M.  \& Riotto, A. 2010c, 
ApJ 717, 526.

\bibitem[Maggiore \& Riotto(2010d)]{MR4} Maggiore, M.  \& Riotto, A. 2010d, 
MNRAS 405, 1244.

\bibitem[Maldacena(2003)]{maldacena}
 Maldacena J.  2003,
JHEP 0305, 013. 

\bibitem[Matarrese et al.(1986)]{MLB} Matarrese, S., 
Lucchin, F., \& Bonometto, S.~A.\ 1986, ApJ., 310, L21. 

\bibitem[Matarrese et~al.(2000)]{MVJ}
  Matarrese, S., Verde, L.  \& Jimenez,~R. 2000,
  ApJ { 541}, 10.

\bibitem[Matarrese \& Verde(2008)]{MV} Matarrese, S. \& Verde, L.  2008, ApJ., 677, L77. 

\bibitem [Matarrese \& Verde(2009)]{MV2009} Matarrese, S.  \& Verde, L. (2009).  
ApJ 706, L91.

\bibitem[{Moscardini et al.}(1991)]{MMLM} Moscardini,.
L.,  Matarrese, S., Lucchin, F., \& Messina, A.\ 1991, MNRAS, 248, 424 

\bibitem[Neistein \& Dekel(2008)]{Neistein}
  Neistein, E. \& Dekel, A. 2008,
  arXiv:0708.1599 [astro-ph].

\bibitem[Peacock \& Heavens(1990)]{PH90}
Peacock, J.A. and Heavens, A.F., 1990, MNRAS 243, 133.

\bibitem[Pillepich {et~al.}(2008)]{PPH}
Pillepich, A. Porciani, C.  \& Hahn, O.  2008,
MNRAS, 402, 191.

\bibitem[Press \& Schechter(1974)]{PS} 
Press,  W. H. \& Schechter,~P. 1974,
ApJ {  187},  425.

\bibitem[Robinson \& Baker (2000)]{RB}
Robinson, J. \& Baker, J.~E. 
MNRAS, 311, 781.

\bibitem[Robinson et al.(2000)]{RGS} Robinson, J.,
Gawiser, E., \& Silk, J.\ 2000, ApJ, 532, 1. 

\bibitem[Shen et al. (2006)]{shen} Shen, J., Abel, T.  Mo, H.  and Sheth, R.K., 
2006,    ApJ.  {645}, 783.

\bibitem[Sheth(1998)]{Sheth1998} Sheth R., MNRAS, 300, 1057

\bibitem[Sheth, Mo \& Tormen (2001)]{SMT} Sheth R., Mo H., Tormen G., 2001, MNRAS, 323, 1.

\bibitem[Sheth \& Tormen(1999)]{ST} Sheth R. \& Tormen G., 1999, MNRAS, 308, 119.

\bibitem[Sheth \& Tormen(2002)]{ST2002} Sheth R. \& Tormen G., 2002, MNRAS, 329, 61.

\bibitem[Slosar  {et~al.}(2008)]{slosar}
Slosar, A., Hirata, C., Seljak, U., Ho, S., Padmanabhan, N. E.    2008,  JCAP 08, 031. 

\bibitem[Smith {et al}(2010)]{zal}
Smith, K.M. , Senatore, L. and  Zaldarriaga, M. 2010, 
JCAP 1001, 028.

\bibitem[Springel et al.(2005)]{Springel:2005nw}
  Springel, V. {\it et al.} 2005,
  Nature {435}, 629

\bibitem[Tinker {et~al.}(2008)]{Tinker:2008ff}
  Tinker J.~L.~{\it et al.} 2008,
  ApJ  688, 709.

\bibitem[Wagner {et ~al.} (2010)]{t}
Wagner, C.  Verde, L.  and Boubekeur, L., 
  arXiv:1006.5793 [astro-ph.CO].



\bibitem[Warren {et~al.}(2006)]{Warren:2005ey}
  Warren, M.~S. et al. 2006,
  ApJ{ 646}  881.

\bibitem[Zentner(2007)]{Zentner}
  Zentner, A.~R. 2007,
  Int.\ J.\ Mod.\ Phys.\  D {16}  763.

\bibitem[Zhang \& Hui(2006)]{Zhang:2005ar}
  Zhang, J.~ \& Hui, L.,
 ApJ  {641},  641.


\end{thebibliography}

\label{lastpage}

\end{document}